\documentclass[aps,prx,twocolumn,amsmath,nofootinbib,amssymb,longbibliography]{revtex4-2}
\usepackage{natbib}
\usepackage{graphicx}
\usepackage{color}
\usepackage[colorlinks, linkcolor=blue, citecolor=red]{hyperref}
\usepackage{amssymb}
\usepackage{gensymb}
\usepackage{float}
\usepackage{amsmath}
\usepackage{tabularx,graphicx}
\usepackage{epstopdf}
\usepackage{latexsym}
\usepackage{color, colortbl}
\usepackage{psfrag}
\usepackage{bbm}
\usepackage{bbold}
\usepackage{bm}
\usepackage{blindtext}
\usepackage{titlesec}
\usepackage{dsfont}
\usepackage{feynmp}
\usepackage{slashed}
\usepackage{multirow}
\usepackage{appendix}
\usepackage{ragged2e}
\usepackage[normalem]{ulem}

\def \tr{{\mbox{tr~}}}

\def \be{\begin{equation}}
\def \ee{\end{equation}}
\def \ba{\begin{array}}
\def \ea{\end{array}}
\def \bea{\begin{eqnarray}}
\def \eea{\end{eqnarray}}
\def \nn{\nonumber}
\def \ve {\varepsilon}
\renewcommand{\vec}[1]{\boldsymbol{#1}}
\def \k {{\vec k}}

\def \bk{{\bf k}}

\def \l{{\lambda}}

\def \a{{\alpha}}

\def \D{{\Delta}}
\def \d{{\delta}}
\def \w{{\omega}}
\def \s{{\sigma}}

\def \k{{\kappa}}

\def \ve{{\varepsilon}}

\def \z{{\zeta}}

\def \ba{\begin{align*}}
\def \ea{\end{align*}}

\newcounter{indice}

\def \mrm{\mathrm}

\def \bs{\boldsymbol}
\def \mc{\mathcal}

\date{~\today}

\begin{document}
\title{The field theory of a superconductor with repulsion }
\author{Amir Dalal}
\affiliation{Department of Physics, Bar-Ilan University, 52900, Ramat Gan, Israel}
\affiliation{Center for Quantum Entanglement Science and Technology, Bar-Ilan University, 52900, Ramat Gan, Israel}
\author{Vladyslav Kozii}
\affiliation{Department of Physics, Carnegie Mellon University, Pittsburgh, Pennsylvania 15213, USA}
\author{Jonathan Ruhman}
\affiliation{Department of Physics, Bar-Ilan University, 52900, Ramat Gan, Israel}
\affiliation{Center for Quantum Entanglement Science and Technology, Bar-Ilan University, 52900, Ramat Gan, Israel}

\begin{abstract}
A superconductor emerges as a condensate of electron pairs, which bind despite their strong Coulomb repulsion. Eliashberg's theory elucidates the mechanisms enabling them to overcome this repulsion and predicts the transition temperature and pairing correlations. However, a comprehensive understanding of how repulsion impacts the phenomenology of the resulting superconductor remains elusive. We present a formalism that addresses this challenge by applying the Hubbard-Stratonovich transformation to an interaction including instantaneous repulsion and retarded attraction.  
We first decompose the interaction into frequency scattering channels and then integrate out the fermions. 
The resulting bosonic action is complex and the saddle point corresponding to Eliashberg's equations generally extends into the complex plane and away from the physical axis. 
{ We numerically determine this saddle point using the gradient descent method, which is particularly well-suited for the case of strong repulsion.}
We then turn to consider fluctuations around this complex saddle point. The matrix controlling fluctuations about the saddle point is found to be a non-Hermitian symmetric matrix, which generally suffers from \textit{exceptional points} that are tuned by different parameters. These exceptional points may influence the thermodynamics of the superconductor. For example, within the quadratic approximation the upper critical field  sharply peaks at a critical value of the repulsion strength related to an exceptional  point appearing at $T_c$. Our work facilitates the mapping between microscopic and phenomenological theories of superconductivity, particularly in the presence of strong repulsion. It has the potential to enhance the accuracy of theoretical predictions for experiments in systems where the pairing mechanism is unknown.
\end{abstract}
\maketitle

\section{Introduction}

The Bardeen-Cooper-Schrifer (BCS) theory~\cite{BCS} gives a microscopic picture of how metals become unstable towards a superconducting state. It is based on the assumption that electronic excitations weakly attract each other when their energy is lower than  the Debye frequency. 
This relatively simple assumption then leads to a theory that offers both important conceptual insight and formidable predictive power.  The theory of Gor'kov~\cite{gor1959microscopic} maps  BCS theory to  a  Ginzburg-Landau (GL) theory, creating a bridge between the microscopic pairing picture and the resulting long-wavelength emergent phenomena, further enhancing the predictive power of BCS theory.

However, BCS theory does not provide a complete picture of the microscopic origin of pairing. In particular, the static interaction between electrons is naively expected to be repulsive, at least within a classical screening theory. 
This naturally leads to the question regarding the quantum origin of the attraction that is assumed in BCS theory.  Morel and Anderson~\cite{Anderson-Morel} used Eliashberg's theory~\cite{eliashberg1960interactions,eliashberg1961temperature,marsiglio2020eliashberg} to show that a pairing instability may occur even when the  interaction is repulsive. The key ingredient that enables the pairing  is retardation of the phonon attraction compared to the instantaneous  Coulomb repulsion. 
Their solution is characterized by frequency dependent pair correlations that change sign between the high and low frequency regimes in a way that exploits the attraction while avoiding the repulsion.
This picture is also amenable within the renormalization group technique, where the effectiveness of the instantaneous repulsion is reduced when dressed with virtual excitations to high energy, while the retarded part is unaffected, thus reducing the repulsion in comparison to the attraction~\cite{ShankarRMP,shankar2011renormalization}. 

Deriving a GL theory that captures the fluctuations around a solution of the Eliashberg equations with a repulsive interaction is, however, not as straightforward as in the case of BCS theory. Nonetheless, it is an important goal, especially for superconductors where Coulomb repulsion is expected to be strong, such as two-dimensional systems~\cite{cao2018unconventional,zhou2021superconductivity,zhou2022isospin,yang2018enhanced,ghazaryan2021unconventional,you2022kohn,tchoumakov2020superconductivity,wagner2023superconductivity}, low-density systems~\cite{Klimin,ruhman2017pairing,gastiasoro2020superconductivity,butch2011superconductivity,behnia2017fragility,prakash2017evidence} and 
possibly even in strongly correlated materials where Eliashberg theory shows unexpected success~\cite{chubukov2020interplay,chowdhury2020unreasonable}.

To better understand the challenge in obtaining the GL theory we may consider  a Hubbard-Stratonovich (HS) transformation~\cite{stratonovich1957method,hubbard1959calculation,altland2010condensed} from the microscopic-fermionic theory to the bosonic one. This is done in two steps. First, the interaction is replaced by a Gaussian integral over a bosonic auxiliary field, which is coupled to a fermion bi-linear. Then the fermions are integrated out to obtain the desired bosonic theory.  

For a simple contact interaction the coupling between the auxiliary bosonic field and the fermions must be real or imaginary, depending on whether the coupling is attractive or repulsive, respectively.
However, a realistic interaction combines instantaneous Coulomb repulsion and retarded phonon attraction~\footnote{Or any other boson that mediates an attractive interaction.}. Thus, an ambiguity arises when performing the HS transformation.

We show that the ambiguity with the HS transformation is signifying a delicate issue regarding the saddle point of the superconducting action  when repulsion is present. Namely, this saddle point may become complex, lying outside the original field-integration path. 
To demonstrate this we first breakdown the interaction matrix into its eigenchannels, which can be in frequency, momentum and spin-orbital space. 
We perform the HS transformation to each eigenchannel separately, such that the repulsive ones are coupled to the fermions via an imaginary coupling, while the attractive ones with a purely real coupling. 
The resulting action is thus a complex functional and, as mentioned above, the saddle point generally lies in the complex plane. 


{ We then obtain the numerical solution of the saddle-point equations using the {\it gradient descent } method~\cite{kantorovich2016functional}. When repulsion is present, this method converges faster than the method of iterating the non-linear Eliashberg equations~\cite{margine2013anisotropic,schrodi2019increased}. Moreover, it is capable of obtaining the solution in the strong repulsion limit, where the iterative approach breakdowns altogether. 
 Finally, we derive the} Ginzburg-Landau theory for the fluctuations about this saddle point in the presence of strong repulsion. Because the saddle point is not necessarily on the physical integration manifold, the expansion about the saddle point is a ``steepest descent''  approximation of the field integral~\cite{bender1999advanced}.

For concreteness, we  apply our theory to the well known Morel-Anderson model of an instantaneous repulsion and retarded attraction~\cite{Anderson-Morel,pimenov2022quantum,pimenov2022twists}. We  first demonstrate the solution of the saddle-point equations  using the gradient descent method and compare the performance to a straightforward iteration technique.  We then show how to incorporate the normal-state self-energy corrections~\cite{Protter2021Functional,chubukovGLtheory}, which are crucial for an accurate description of the superconducting state.

Next, we discuss the derivation of a GL theory for the fluctuations about the saddle-point solution. Within a quadratic expansion, the fluctuations of  different eigenmodes are generally coupled through a non-Hermitian symmetric matrix, which may have complex eigenvalues. In particular, the eigenvalues of the matrix generically incur exceptional points.  These are the points at which the spectrum of the matrix becomes degenerate and the matrix itself is defective in a sense that it is non-diagonalizable~\cite{kato1966analytic,doi:10.1126/science.aar7709}. They only appear in the presence of repulsion and can be tuned by different parameters of the system, such as temperature, undulation wavelength, and coupling strength. 
Interestingly, we find that the temperature of the exceptional point in the lowest eigenvalue branch is always higher than the transition temperature, except for a critical value of the repulsion strength where the two temperatures are equal.
At this value of the repulsion strength the fluctuation matrix is defective at $T_c$. The properties of such an {\textit {exceptional superconductor}} remain to be uncovered. 
Finally, we use the quadratic approximation to compute upper critical field $H_{c2}\propto (1-T/T_c)/\xi_{GL}^2$ close to $T_c$. 
The Ginzburg-Landau coherence length $\xi_{GL}$ is found to be strongly diminished near the critical repulsion strength where the exceptional point appears at $T_c$. Away from the critical repulsion strength $\xi_{GL}$ depends monotonically on repulsion in a way that depends on the details of the interaction and generally deviates from the Gor'kov-BCS result~\cite{gor1959microscopic}, $\xi_{GL}^{BCS} =  \sqrt{7 \zeta(3)} v_F / 4\pi\sqrt{3} T_c $, where $v_F$ is the Fermi velocity. 

Our results are expected to be important for any inclusive study of superconductivity from the weak coupling perspective, especially in low-density and two-dimensional systems. Furthermore, our theory  may also contribute to the efficiency of numerical solvers of the non-linear Eliashberg equation where strong repulsion is included. 
Finally, we comment that it may also be relevant to the understanding of the Kohn-Luttinger~\cite{KohLuttinger} mechanism of superconductivity, anytime the system lacks rotational symmetries and repulsive and attractive channels mix.

The rest of this paper is organized as follows. First we briefly review some of the properties  of Eliashberg theory essential to our paper.  In Section \ref{sec:The superconducting action in the presence of a repulsive interaction} we describe the eigenchannel decomposition of the interaction and perform the Hubbard-Stratonovich transformation. In Section \ref{sec:The Saddle point solution} we numerically obtain the complex saddle point solution of the Hubbard-Stratonovich action, show its equivalence to the solution of the Eliashberg equation, and discuss its dependence on repulsion. In Section \ref{sec:SE} we show how to include the normal state self-energy corrections and discuss their influence on the saddle-point solution. Finally, in  Section \ref{sec:GL theory} we derive the long-wavelength theory for fluctuations around the  saddle point and use it to compute the influence of the repulsion strength on the upper critical field close to $T_c$.

\subsection{Brief review of Eliashberg and Morel-Anderson theory}
Let us quickly review some of the essential properties of Eliashberg theory, before describing how to incorporate it in a field theoretic formalism. We will mainly focus on a simplified model, where the interaction between electrons in the $s$-wave channel includes an instantaneous Coulomb repulsion and a retarded attraction (see, for example, Refs.~\cite{Anderson-Morel,pimenov2022quantum})
\be\label{Eq:V_AM}
\hat V_{\w,\w'} = {\lambda\over N_F}\left[\mu -{\omega_D^2 \over (\w-\w')^2+ \w_D^2}\right]\,,
\ee
where $\w_D$ is the frequency of an Einstein phonon mode that mediates the attraction, 
$N_F$ is the fermionic density of states at the Fermi level, and the dimensionless parameters $\lambda$ and $\mu$ quantify the total coupling strength and the relative strength of the repulsion, respectively. The quantity
$\lambda\mu = \langle q_{TF}^2  /2(q^2+q_{TF}^2)\rangle_{FS}$ is naively assumed to be the Fermi-surface average  over the screened Coulomb interaction~\cite{margine2013anisotropic}, where $q_{TF}$ is the Thomas-Fermi screening length. 
In the case $\mu>1$ the bare interaction is repulsive at all frequencies. We emphasize that as long as only classical screening is taken into account we expect $\mu$ will always be larger than unity. Moreover, $\mu>1$ even within the random-phase approximation (RPA) when projecting to the $s$-wave channel. 

For the simplified model \eqref{Eq:V_AM} Eliashberg's equations become momentum independent 
\begin{align}  \label{eq:Eliashberg}
\begin{split}
    &\Delta(\omega)=-\frac{\pi T N_F}{2} \sum_{\omega'}\frac{\hat{V}_{\w,\w'}\Delta(\omega')}{\sqrt{[\omega'+i\Sigma(\omega')]^2+|\Delta(\omega')|^2}}\\
    &\Sigma(\omega)=\frac{\pi T N_F}{2} \sum_{\omega'}\frac{\hat{V}_{\w,\w'}[i\omega'-\Sigma(\omega')]}{\sqrt{[\omega'+i\Sigma(\omega')]^2+|\Delta(\omega')|^2}}\,,
\end{split}
\end{align}
where $\omega = 2\pi T(n+1/2)$ is a fermionic Matsubara frequency, $T$ is the temperature and $N_F$ is the density of states at the Fermi level.
$\Sigma(\omega)$ is the normal state self-energy, which is  purely imaginary and anti-symmetric $\Sigma(\omega) = -\Sigma(-\omega)$. Although this term is sometimes neglected, it significantly affects the superconducting properties including the transition temperature, especially in the presence of Coulomb repulsion. 
The self-consistent equation for the pairing field $\Delta(\omega)$, obeying the symmetry $\D(\w) = \D(-\w)$ because we implicitly assumed singlet pairing without any momentum dependence.  

In their more general form, Eqs.~(\ref{eq:Eliashberg})  are the foundation of our most advanced microscopic understanding of superconductivity. They capture spectral properties that go beyond BCS theory~\cite{SchriefferScalapinoWilkins} and are the workhorse of quantitative calculations for conventional and unconventional superconductors~\cite{margine2013anisotropic,sanna2018ab}. They are also capable of capturing non-Fermi liquid behavior in strongly correlated systems and its interplay with quantum criticality and superconductivity~\cite{abanov2003quantum,chowdhury2020unreasonable}. 

One of the most important hallmarks of these equations is that they support a non-trivial solution even when the interaction in Eq.~\eqref{Eq:V_AM} is positive (repulsive) at all frequencies ($\mu>1$). This solution is characterized by a sign-changing gap function of the form
\be \label{eq:Eliashber_sol}
\Delta(\omega) = \begin{cases}
\Delta_0, &\w\ll \w_D  \\
-\Delta_1, &\w\gg  \w_D,
\end{cases}
\ee
where the ratio $\Delta_1/\D_0$ is positive. Morel and Anderson~\cite{Anderson-Morel} approximated the interaction from Eq.~\eqref{Eq:V_AM} using a step function and found that $\D_1/\D_0=1/(\lambda/\mu_*-1)$ and $T_c\sim \w_D\exp[-1/(\lambda-\mu_*)]$, where $\mu_* =  \mu /(\lambda^{-1}+ \mu \ln \epsilon_F/\w_D)$.



An intuitive understanding of the sign change  is obtained by making an analogy between  frequency   and  momentum dependence of the gap. In particular, one may consider interactions with multiple angular-momentum scattering channels, where the strongest interaction is in the repulsive $s$-wave channel, in addition to some weaker attractive channel with higher angular momentum (which is the case in the Kohn-Luttinger mechanism~\cite{KohLuttinger}). Clearly, because the $s$-wave is nodeless, the higher angular momentum channel must have nodes to establish orthogonality. The role of the node in the frequency-dependent gap function is similar. We can imagine decomposing the frequency-dependent interaction \eqref{Eq:V_AM} into  scattering channels, where the repulsive part is nodeless and therefore the attractive channel forming the superconducting state must have nodes. 
 
However, obtaining an unbiased (and numerically exact) solution for any value of $\mu$ can be challenging. For example, the performance of an iterative method deteriorates with increasing repulsion strength, $\mu$,  which eventually breaks down at some critical value of $\mu$.

Moreover, an important question regards the microscopic derivation of a Ginzburg-Landau theory for the fluctuations around this solution.
Such a theory is important  when the proposed pairing interaction  includes strong repulsion~\cite{abanov2003quantum,ruhman2016superconductivity,ruhman2017pairing,gastiasoro2020superconductivity,tchoumakov2020superconductivity,kozii2019superconductivity,lewandowski2021pairing,klein2019multiple}. For example, the Ginzburg-Landau theory can be qualitatively different when the pairing interaction is long-ranged~\cite{yang2000low}.
Thus, developing a formalism to tackle these problems is important for a wide scope of problems, which go beyond the specific model in Eq.~\eqref{Eq:V_AM}.

\section{The superconducting action in the presence of a repulsive interaction} \label{sec:The superconducting action in the presence of a repulsive interaction}
In this section, we present the methodology for developing the field theory for a superconducting state stemming from a pairing interaction with repulsion. To this end, we use the well known HS transformation~\cite{altland2010condensed}. However, before performing the transformation  we need to distinguish between the attractive and repulsive channels. Therefore, we first discuss the decomposition of the interaction in Eq.~\eqref{Eq:V_AM}  into scattering channels in frequency and momentum space.

\subsection{Frequency and Momentum Eigen-channels of the Pairing Interaction}
To demonstrate the decomposition of the interaction into channels we consider a generic pairing interaction
\begin{equation}\label{Eq:S_I}
    \mc S_{I}=\frac{T}{L^3}\sum_{k,p,Q}{\Lambda}^\dag_k(Q)\hat{V}_{k,p}\Lambda_p(Q)\,,
\end{equation}
where $k,p$ are four-vectors, $k=\{\omega,\vec{k}\}$, and $\hat V_{k,p}$ is a generic interaction which is independent on the center of mass coordinate $Q$. The Cooper-pair bilinears ${\Lambda}^\dag _k(Q)$ and $\Lambda_p(Q)$ are given by
\begin{align}
    &\Lambda_p(Q)=\psi_{-p,\downarrow}\psi_{p+Q,\uparrow}\nonumber \\
    &{\Lambda}_k^\dag(Q)=\psi^{\dagger}_{k+Q,\uparrow}\psi^\dagger_{-k,\downarrow}\,,
    \nonumber
\end{align}
Note that here we have explicitly assumed singlet pairing only. 

The first step in the HS transformation is to replace the interaction in Eq.~\eqref{Eq:S_I} with a Gaussian path integral over an auxiliary field, which couples linearly to $\Lambda_p (Q)$. 
The Gaussian integral must be convergent and consequently the coupling is either purely real or purely imaginary for an attractive or repulsive interaction, respectively. However, in the general case the interaction can not be defined as purely attractive or purely repulsive. For example, the interaction in  Eq.~\eqref{Eq:V_AM} contains both repulsive and attractive components. This raises the question about the correct method to perform such a HS transformation.  

\begin{figure}[h!]
\centering\includegraphics[width=0.45\textwidth]{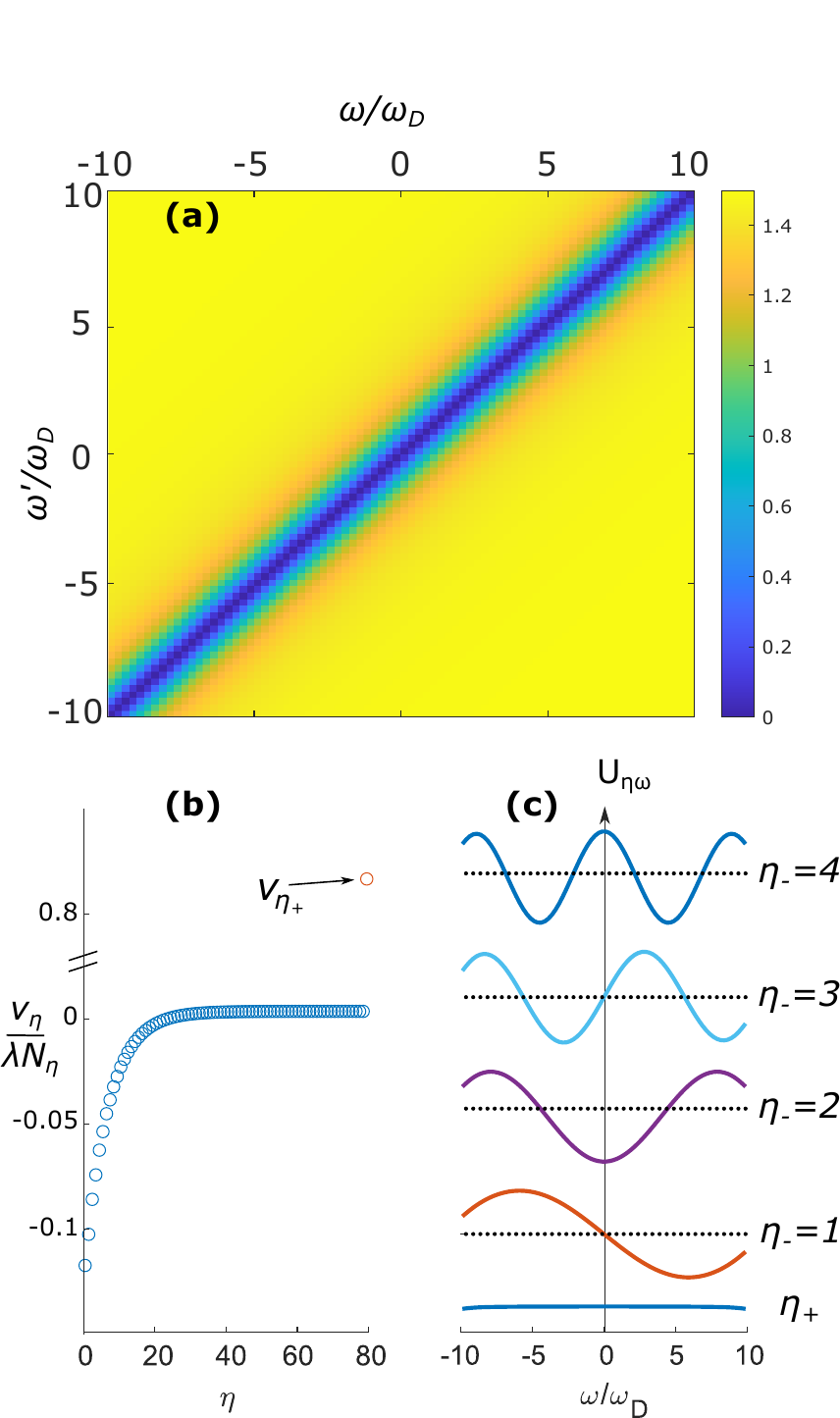}
    \caption{Eigenvalue decomposition of the Anderson-Morel interaction (Eq. \ref{Eq:V_AM}). \textbf{(a)}. The interaction in $\omega-\omega'$ plane for values of $\mu=1$ and $\lambda=1.5$.  \textbf{(b)}. The different eigenvalues $v_{\eta}$ normlaized by $\lambda$ and the number of eigenvalues $N_\eta$. Note that all of them are negative (i.e., attractive), except for a single positive one, which corresponds to a repulsive channel $v_{\eta_+}$.  \textbf{(c)} The eigenvectors $U_{\eta,\omega}$  for the first 4 attractive $\eta_-$ and the single repulsive $\eta_+$  as a function of $\omega$.}
    \label{fig:EigLorentz}
\end{figure}

To answer this question we decompose the interaction into its eigen channels in $k$-space
\begin{align}\hat{V}_{k,p}=\sum_{\eta}v_{\eta}{U}^*_{\eta,k}U_{\eta,p}\,,
    \label{eq:Intkp}
\end{align}
where $\eta$ labels different orthogonal channels, $v_{\eta}$ are the eigenvalues and $U_{\eta,p}$ are the eigenvectors, such that $\sum_{\eta}{U}^*_{\eta,k}U_{\eta,p}=\delta_{k,p}$ and $\sum_{k}{U}^*_{\eta,k}U_{\eta',k}=\delta_{\eta, \eta'}$.
The eigenvectors $U_{\eta,k}$ define the scattering channels for which the interaction is diagonal 
\begin{align}\label{eq:phi_transformation}
    \varphi_{\eta}(Q)=\sqrt{\frac{T}{L^3}}\sum_{k}U_{\eta,k}\Lambda_k(Q)\,,
\end{align}
The  interaction in Eq.~\eqref{Eq:S_I} then assumes the simple form
\begin{align}\label{eq:eigs_decomp}
    \mc S_I=\sum_{Q,\eta} v_\eta {\varphi}^\dag_{\eta}(Q) \varphi_{\eta}(Q).
\end{align}

A general interaction will have both positive and negative eigenvalues. In fact, due to Coulomb repulsion this is always the case for electrons. Throughout the paper we will refer to eigenchannels corresponding to positive eigenvalues as ``repulsive'' and eigenchannels corresponding to negative eigenvalues as ``attractive'', as follows 
\begin{align}
\begin{cases}
v_{\eta}>0 & \mrm{repuslive},\\ 
v_{\eta}<0 & \mrm{attractive}.
\end{cases}
\end{align}
We note that a similar mixture of repulsive and attractive
orbital channels was recently considered in Ref. \cite{Henrik2022Constrained}.
Moreover, we will sometimes need to distinguish these two using a $\pm$ subscript notation $\eta_+$ and $\eta_-$, {such that $v_{\eta-}<0$ and $v_{\eta+}>0$}. 
The interaction can then be divided into ``repulsive'' and ``attractive'' parts: 
\begin{align}\label{Eq:S_I eigenbasis form}
&\mc S_I=\mc S_{+}+\mc S_{-}
\\&=\sum_{Q,\eta_+}|v_{\eta_+}| {\varphi}_{\eta_+}^\dag(Q) \varphi_{\eta_+}(Q)-\sum_{Q,\eta_-} |v_{\eta_-}| {\varphi}_{\eta_-}^\dag(Q) \varphi_{\eta_-}(Q).\nn
\end{align}

With Eq.~\eqref{Eq:S_I eigenbasis form} in hand we are in good position to perform the HS transformation (see Section \ref{Sec:HS transformation}). However, let us first briefly review the properties of the eigensystem for the case of Eq.~\eqref{Eq:V_AM} and then make some more general remarks.  
In Fig.~\ref{fig:EigLorentz}(a) we plot the interaction matrix $ \hat V_{\omega,\omega'} = \hat V(\omega-\omega')$ for $\mu=1$, $\lambda=1.5$, $2\pi T=0.25\omega_D$, and cutoff at $10 \w_D$. 
In panel (b) we plot the eigenvalues $v_\eta$ vs. $\eta$. Note that all eigenvalues are negative (i.e., attractive), except a single one, which is positive (i.e., repulsive), denoted by $v_{\eta_+}$.
Lastly, in panel (c), we plot five eigenvectors $U_{\eta,\omega}$ with the  largest absolute value eigenvalues as a function of Matsubara frequency, $\omega$. Note that the eigenvectors have well defined parity with respect to $\omega\to-\omega$. These correspond to odd and even frequency channels. In the case of singlet superconductivity, as considered here, the odd-frequency channels do not contribute and their weight in the gap function must be zero. 

Before proceeding to perform the HS transformation we first make a few important remarks. 
First, we note that the eigenvalue decomposition in Eq.~\eqref{eq:Intkp} is well known and used in different contexts. 
For example, in scattering theory a spherically symmetric interaction is decomposed into angular momentum channels, then $\eta$ would correspond to the angular momentum quantum numbers $l,m$ and the eigenvectors $U$ to the spherical harmonics. When a symmetry is present pairing channels belonging to different irreducible representations (irreps) of the symmetry group will decouple. The saddle point for each irrep can then be solved separately. This is the case in the well known Kohn-Luttinger problem~\cite{KohLuttinger}, where full rotational symmetry  is present and $T_c$ is set by the largest attractive channel, regardless of how strong the repulsive ones are. 

However, in many cases repulsive and attractive channels do not belong to different irreps and do not decouple at the saddle point.
Such is the case in Eq.~\eqref{Eq:V_AM}, where the only existing symmetry is time-reversal symmetry. 
This symmetry decouples the odd-frequency and even-frequency channels.  
As we will see, the odd-frequency channels can affect superconductivity in the singlet superconductivity  through the normal state self-energy. 
Another example, where repulsive and attractive channels mix would the Kohn-Luttinger mechanism in a system where the full rotational symmetry is broken (e.g. due to a lattice). 

\subsection{The Hubbard-Stratonovich transformation and the resulting bosonic action}\label{Sec:HS transformation}
We now turn to perform the HS transformation~\cite{altland2010condensed,yang2000low}. As explained, the transformation involves the introduction of a Gaussian integral over an auxiliary field that is linearly coupled to the pairing fields.
However, in order for the Gaussian integrals to be convergent the auxiliary fields can not couple in the same way to the attractive and repulsive channels in Eq.~\eqref{Eq:S_I eigenbasis form}. Namely, the repulsive/attractive channels must be coupled by purely imaginary/real couplings:
\begin{widetext}
\begin{align}\label{eq:HS_transfrmation}
 &e^{-\mc S_{-}}=\exp \left[\sum_{Q,\eta_-}{|v_{\eta_-}| {\varphi}_{\eta_-}^\dag\varphi_{\eta_-}}\right]
    =\int \mathcal D [f^{*}_{\eta_-},f_{\eta_-}]\exp\left[{-\sum_{Q,\eta_- }\frac{|f_{\eta_-}|^2}{|v_{\eta_-}|}+\sum_{Q,\eta_-}\left(f_{\eta_-}{\varphi}_{\eta_-}^\dag+f^{*}_{\eta_-} \varphi_{\eta_-}\right)}\right],
  \\
 &e^{-\mc S_{+}}=\exp\left[{-\sum_{Q,\eta_+}|v_{\eta_+}| {\varphi}_{\eta_+}^\dag\varphi_{\eta_+}}\right]
    =\int \mathcal D[f^{*}_{\eta_+},f_{\eta_+}]\exp\left[{-\sum_{Q,\eta_+}\frac{|f_{\eta_+}|^2}{|v_{\eta_+}|}+i\sum_{Q,\eta_+}\left(f_{\eta_+}{\varphi}_{\eta_+}^\dag+f^{*}_{\eta_+} \varphi_{\eta_+}\right)}\right],\nn
\end{align}
where we introduced the HS auxiliary fields $f_\eta(Q)$ and suppressed index $Q$ for brevity.  
Using the equations above, the interaction in Eq.~\eqref{Eq:S_I} becomes 
\begin{align}\label{eq:S_I ater HS}
\mc S_{I}\to \sum_{Q,\eta}\left[\frac{|f_{\eta}(Q)|^2}{|v_{\eta}|}-\z_\eta f_{\eta}(Q){\varphi}_{\eta}^\dag(Q)-\z_\eta f^{*}_{\eta}(Q) \varphi_{\eta}(Q)\right] = \sum_{Q,\eta} \frac{|f_{\eta}(Q)|^2}{|v_{\eta}|}-\sum_{k, Q}\left(\Lambda^\dagger_k(Q) \Delta_{1,k}(Q)+\Lambda_k(Q) \Delta^*_{2,k}(Q)\right),
\end{align}
\end{widetext}
where
\[
\zeta_\eta=\begin{cases}
1 & v_{\eta}<0\\ 
i & v_{\eta}>0
\end{cases} \,,
\]
and 
\begin{align} \label{eq: D1 and D2 def}
\Delta_{1,k} (Q)&\equiv\sqrt{\frac{T}{L^3}}\sum_{\eta}\zeta_{\eta} U^*_{\eta,k} f_{\eta} (Q) \\
\Delta^*_{2,k}(Q)&\equiv\sqrt{\frac{T}{L^3}}\sum_{\eta}\zeta_{\eta} U_{\eta,k} f^*_{\eta}(Q)\,.\nn
\end{align}
The last equality on the RHS of Eq.~\eqref{eq:S_I ater HS} was obtained using the relation \eqref{eq:phi_transformation}. It is important to note that in the presence of repulsion the HS fields, $\Delta_{1,k}$ and $\Delta^{*}_{2,k}$, are not complex conjugates. 

Using these notations, the full action assumes the form 
\begin{align} \label{Eq:SHS}
\begin{split}
\mc S_{HS}=\sum_{Q,\eta}\frac{|f_{\eta}(Q)|^2}{|v_{\eta}|}
+\sum_{k,Q}\Psi^{\dagger}_{k+Q} \,\mathcal G_{k}^{-1}(Q)\,
\Psi_k\,,
\end{split}
\end{align}
where $ \Psi_k^\dag = ( \psi_{k\uparrow}^\dag,\psi_{-k\downarrow})$ 
is the Nambu spinor, $G_0(k)=(-i\omega+\xi_k)^{-1}$ is the bare Green's function and the Gor'kov Green's function is defined by
\begin{align}
    \mathcal G_{k}^{-1}(Q)= 
    \begin{pmatrix}
G_0^{-1}(k) \delta_{Q,0}  & -\Delta_{1,k}(Q) \\
-\Delta^{*}_{2,k{ + Q}}({ -}Q) & -G_0^{-1}(-k)\delta_{Q,0}
\end{pmatrix}.
\end{align}


Finally, we perform the last step in the HS transformation, the integration over the fermionic fields. This yields a fully bosonic action 
\begin{align}\label{eq:GL_action}
\begin{split}
    \mc S_{HS}=\sum_{Q,\eta}\frac{|f_{\eta}(Q)|^2}{|v_{\eta}|}-\mathrm{tr}\ln{\mathcal G^{-1}_{k} (Q)}.
    \end{split}
\end{align}

In what follows we will show that  when repulsive channels are present, the saddle point of this action lies outside the integration region of the fields $(f_\eta^*,f_\eta)$ introduced by  Eqs.~\eqref{eq:HS_transfrmation}.  
We also note that because  $\Delta_{1,k}$ and $\Delta^{*}_{2,k}$, are not complex conjugates of one another the Green's function matrix in the $\mrm{tr}\ln$ is not Hermitian.

\begin{figure*}
     \centering
         \includegraphics[width=1\textwidth]{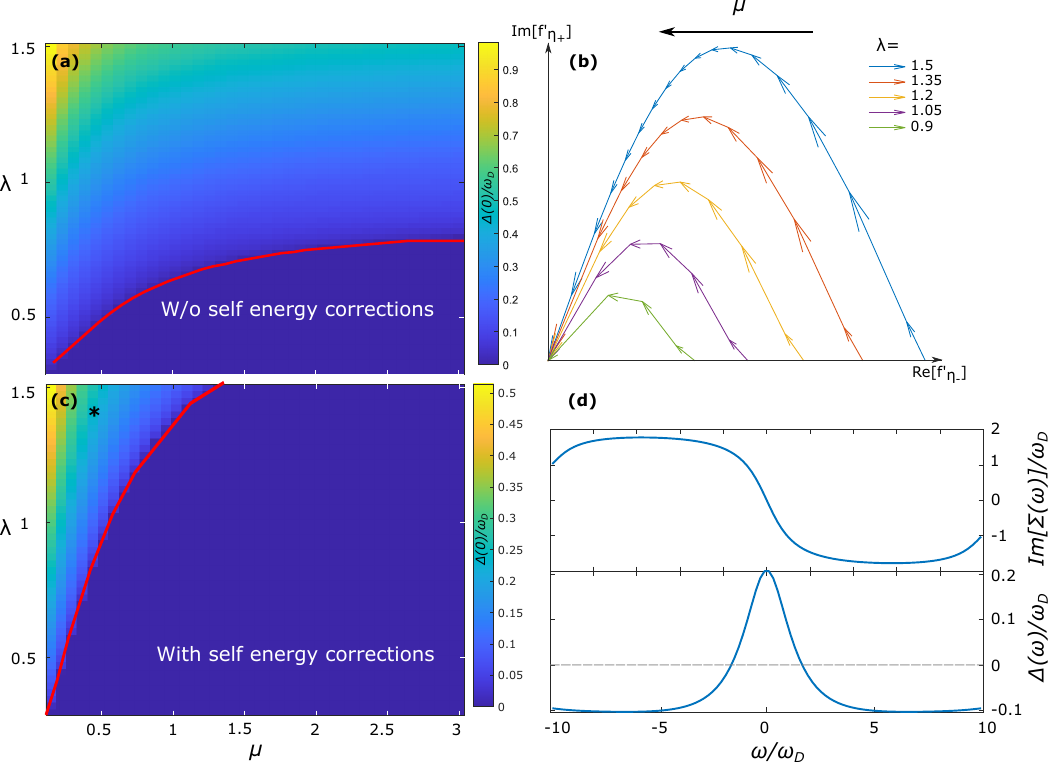}
         \caption{The solution of the saddle-point equations at zero frequency, $\Delta(\omega=0)$, for different values of $\mu$ and $\lambda$, obtained by using the gradient descent procedure described in Sec. \ref{sec:GD}. The temperature is $2\pi T=0.03\omega_D$. \textbf{(a)}, \textbf{(c)}  The saddle-point solution,  $|\Delta(\omega=0)|$, computed without and with self-energy corrections, respectively. \textbf{(b)} The values of the only repulsive channel $\mrm{Im}[f'_{\eta_+}]$ as a function of the largest attractive channel  $\mrm{Re}[f'_{\eta_-}]$ at the saddle-point solution for different $\lambda$ (indicated in the legend) and $\mu$. The arrows indicate the flow direction with increasing $\mu$, starting from zero. \textbf{(d)}  The functions $\Sigma(\w)$ (top panel)  and $\D(\w)$ (bottom panel) at the  saddle-point solution computed for  $\lambda=1.3$ and $\mu=0.5$, which is marked by the asterisk in panel (c). }
         \label{fig:GapWdVsMuGD}. 
\end{figure*}

\section{The saddle-point solution}
\label{sec:The Saddle point solution}
After obtaining the field theory in Eq.~\eqref{eq:GL_action} we turn to explore the properties of the saddle point. We will also show that the solution of this saddle point   satisfies Eliashberg's pairing equation. 

For convenience we transform the fields $( f_\eta^*,f_\eta)$ to  a real representation
\begin{align}\label{eq:f' and f'' to f and fbar}
\begin{split}
        f_{\eta}&= f'_{\eta}+if''_{\eta}\\
        f^*_{\eta}&= f'_{\eta}-if''_{\eta} \,.
    \end{split}
\end{align}
{We note that the integration contour in Eq.~\eqref{eq:HS_transfrmation} implies that  both  $f'_\eta$ and $f''_\eta$ are real fields covering the whole real space, $\mathbb{R}^{N_\eta}\otimes \mathbb{R}^{N_\eta}$, where $N_\eta$ is the number of fields. }

To obtain the saddle point, we take the  derivatives of the action in Eq.~\eqref{eq:GL_action} with respect to $f'_{\eta}(0)$, $f''_{\eta}(0)$ at $Q=0$, which yields~\footnote{By taking the derivative with respect to the fields at $Q=0$ we have restricted our search to states that are spatially homogeneous. In the general case, especially when the interaction is momentum-dependent, one must verify that there are no other saddle-point solutions at finite $Q$, corresponding to Fulde-Ferrell-Larkin-Ovchinnikov or density waves  states.} 
\begin{align}\label{eq:FFbar before}
\begin{split}
     \frac{2f'_{\eta}}{|v_\eta|}&={ -}\sqrt{\frac{T}{L^3}}\rm{tr}\left[\mathcal{G}_{k}(0)
 \begin{pmatrix}
 0  & \zeta_{\eta} U^*_{\eta,k} \\
 \zeta_{\eta} U_{\eta,k} & 0 
 \end{pmatrix}\right],\\
    \frac{2f''_{\eta}}{|v_\eta|}&={ -}\sqrt{\frac{T}{L^3}}\rm{tr}\left[\mathcal{G}_{k}(0)
 \begin{pmatrix}
 0  & i\zeta_{\eta} U^*_{\eta,k} \\
 -i\zeta_{\eta} U_{\eta,k} & 0 
 \end{pmatrix}\right].
\end{split}
\end{align}

For concreteness, let us focus on the case of Eq.~\eqref{Eq:V_AM}, where the interaction is momentum-independent and the eigenvectors are only functions of frequency. Then we can integrate over momentum and obtain 
\begin{align}\label{eq:saddle point f' and f''}
\begin{split}
           {f'_\eta\over \sqrt{L^3 T}} &=\frac{\pi  N_F |v_\eta| \zeta_\eta}{2} \sum_{\omega} {\frac{\Delta_{1,\omega} U_{\eta,\omega}+\bar\Delta_{2,\omega} U^*_{\eta,\omega}}{\sqrt{\omega^2+\Delta_{1,\omega}\bar\Delta_{2,\omega}}}},\\   
           {f''_\eta\over \sqrt{L^3 T}} &=-i\frac{\pi  N_F|v_\eta| \zeta_\eta}{2} \sum_{\omega} \frac{\Delta_{1,\omega} U_{\eta,\omega}-\bar\Delta_{2,\omega} U^*_{\eta,\omega}}{\sqrt{\omega^2+\Delta_{1,\omega}\bar \Delta_{2,\omega}}} \,.
\end{split}
\end{align}
Notice that we have introduced a new field $\bar \D_2$ instead of $\D_2^*$. 
To understand this we recall that when the interaction has repulsive eigenvalues, $\Delta_1$ is not related to $\Delta_2^*$ by complex conjugation. Consequently, the  saddle-point solution of Eqs.~\eqref{eq:saddle point f' and f''} is only obtained with {\it complex} $f'_{\eta}$ and $f''_{\eta}$. 
In other words, when repulsion is present, the saddle point is not located on the  original field-integration manifold but extended into the complex space $\mathbb{C}^{N_\eta}\otimes \mathbb{C}^{N_\eta}$. 
Therefore, we can no longer identify the fields in Eq.~\eqref{eq:f' and f'' to f and fbar} as complex conjugates of one another. To emphasize this we henceforth distinguish between the asterisk notation, $(.)^*$, which denotes complex conjugation, and the ``bar'' notation $\bar{(.)}$, which defines an independent field $\bar f_\eta$, on pare with $f_\eta$. In particular, we modify our notation to
\begin{align}\label{eq:f* to bar f}
\begin{split}
&f_\eta^* \;\;\to\;\; \bar f_\eta  \equiv f'_\eta - i f_\eta '' \ne f_\eta^*, \\ 
&\D_2^* \;\;\to\;\; \bar \Delta_2  \equiv  \sqrt{\frac{T}{L^3}}\sum_{\eta}\zeta_{\eta} U_{\eta,k} \bar f_{\eta}\ne \D_2^*\,,
\end{split}
\end{align}
while $f_\eta$ and $\D_1$ are still defined as they appear in Eq.~\eqref{eq:f' and f'' to f and fbar} and Eq.~\eqref{eq: D1 and D2 def}, respectively. 

In what follows it will be useful to write the self-consistency equations~\eqref{eq:saddle point f' and f''} in terms of $f_\eta$ and $\bar f_\eta$. 
To that end, we take the sum and difference of Eqs.~\eqref{eq:saddle point f' and f''} with the appropriate coefficients to give the complex representation 
\begin{align}
\label{eq:gapFFbar}
\begin{split}
           {f_\eta \over \sqrt{L^3 T}} &=\pi N_F |v_\eta| \zeta_{\eta}\sum_{\omega} {\frac{\Delta_{1,\omega} U_{\eta,\omega}}{\sqrt{\omega^2+\Delta_{1,\omega}\bar \Delta_{2,\omega}}}}\,,\\                
           {\bar f_\eta \over \sqrt{L^3 T}}&=\pi  N_F|v_\eta| \zeta_{\eta}\sum_{\omega} {\frac{\bar \Delta_{2,\omega}U^*_{\eta,\omega}}{\sqrt{\omega^2+\Delta_{1,\omega}\bar \Delta_{2,\omega}}}}\,, \\
\end{split}
\end{align}
where again $\bar f_\eta = f'_\eta - i f''_\eta$ is now not necessarily the complex conjugate of $f_\eta$.

It is important to note that  the complex saddle point represented by these equations still captures the low-energy physics of the superconductor despite the fact that it is not in the original integration space. This is justified by deforming the integration path in Eqs.~\eqref{eq:HS_transfrmation} to go through the saddle point given by Eqs.~\eqref{eq:saddle point f' and f''} and along the direction of ``steepest descent''~\cite{bender1999advanced}. In this case the Gaussian fluctuations along the path and near the saddle point dominate the low-energy physics of the superconductor.

\subsection{Equivalence to Eliashberg's equation \label{Sec:Eleq}}
Before demonstrating the usefulness of Eqs.~\eqref{eq:saddle point f' and f''} and \eqref{eq:gapFFbar}, we first show that they coincide with Eliashberg's pairing equation.
We multiply both sides of Eq.~\eqref{eq:gapFFbar} by the factor $\zeta_\eta U_{\eta,\omega'}$, sum over all $\eta$ and use Eq.~\eqref{eq:f* to bar f} to obtain
\begin{align}\label{eq:D1 and D2}
\begin{split}  
    \Delta_{1,\omega'}&=-\pi TN_F
\sum_{\omega}\frac{\hat{V}_{\omega',\omega}\Delta_{1,\omega}}{\sqrt{\omega^2+\Delta_{1,\omega}\bar\Delta_{2,\omega}}},\\
    \bar\Delta_{2,\omega'}&=-\pi TN_F\sum_{\omega}\frac{\hat{V}_{\omega',\omega}\bar\Delta_{2,\omega}}{\sqrt{\omega^2+\Delta_{1,\omega}\bar\Delta_{2,\omega}}}.
    \end{split}
\end{align}
{These equations and their solution are identical to Eliashberg's equation, while generalization for the case of momentum-dependent interaction and/or gap functions is obvious.} However, as we will see in Section ~\ref{sec:GD}, from the numerical perspective there is a significant  advantage in solving the equations in the eigenbasis of Eq.~\eqref{eq:gapFFbar}.

\subsection{Numerical saddle-point solution with strong repulsion } 
\label{sec:GD}
The form of the non-linear Eliashberg's equations in Eq.~\eqref{eq:Eliashberg}  [or in Eq.~\eqref{eq:D1 and D2}] is convenient for numerical solution by the method of self-consistent iteration~\cite{margine2013anisotropic,schrodi2019increased}. However, when strong repulsion is present, this method may exhibit numerical instability. For example, the solution tends to oscillate between negative and positive solutions. These instabilities can be somewhat mitigated by updating the gap locally instead of globally or by using a cleaver initial ansatz. 
In this section we { demonstrate the use of the gradient descent method~\cite{kantorovich2016functional} on}
 Eqs.~\eqref{eq:gapFFbar}   to  obtain a stable numerical solution at any $\mu$.

\begin{figure}
    \centering
    \includegraphics[width=0.4\textwidth]
    {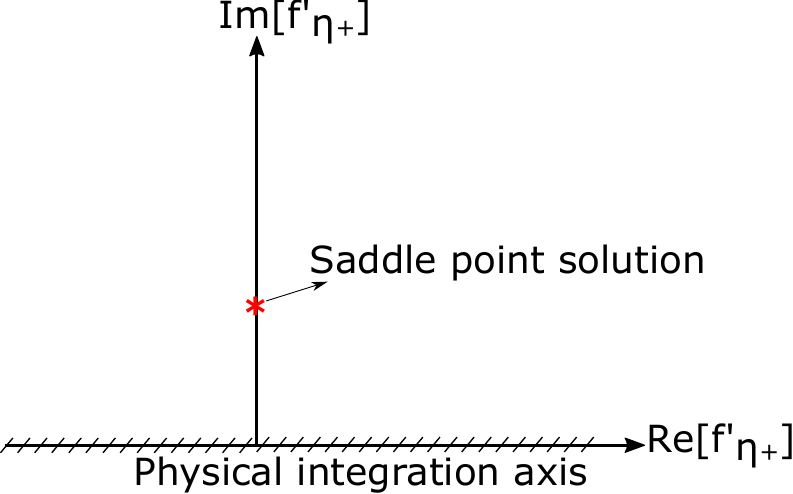}
    \caption{ The schematic location of the saddle-point solution in the complex plane of the field $f_{\eta_+}'$, which is associated with the repulsive eigenvalue $v_{\eta_+}$. Before extending this field into the complex plane, i.e., on the physical integration axis, it took real values $f_{\eta_+}' \in (-\infty,\infty)$, as defined in Eq.~\eqref{eq:f' and f'' to f and fbar}. At the saddle point, however, $\mrm{Re}[f'_{\eta_+}]=0$, so $f'_{\eta_+}$ is purely imaginary, and $f_{\eta_+}'' = 0$.  }
    \label{fig:GradientPassing}
\end{figure}

\begin{figure}
    \centering
\includegraphics[width=0.47\textwidth]
    {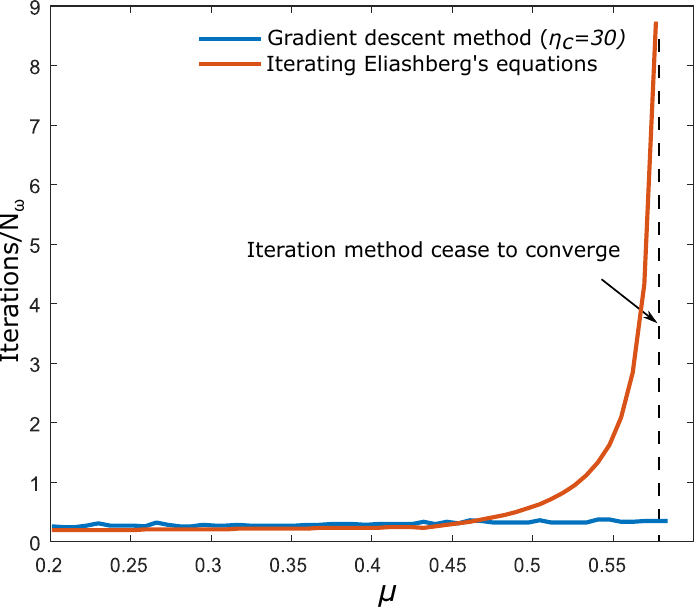}
    \caption{The number of iterations, normalized by the number of Matsubara  frequencies, needed for the Eliashberg solution to converge as a function of repulsion strength $\mu$ for $2\pi T/\w_D =0.03$ and $\lambda = 1$.
    The blue and red curves correspond to different solution methods, eigenvalue decomposition and regular iteration of Eliashberg's equation, respectively.
    Convergence is defined by the deviation of less than 1\% of the last iteration's solution. The dashed line in the figure corresponds to the critical value of $\mu$ above which the iteration of Eliashberg's equation ceases to converge to a solution.} 
    \label{fig:Convergent}
\end{figure}

{
To implement the gradient descent method we evolve the fields $f_\eta$ and $\bar f_\eta$ in small increments along the direction at which the action changes most rapidly in the complex space of fields 
\begin{align}\label{eq:grad_passing_update}
\begin{split}
&f^{i+1}_{\eta}=f^i_{\eta}-e_{\eta_+} |v_{\eta}| \frac{\partial \mc S_{HS}}{\partial f^i_{\eta}},\\
    &\bar f^{i+1}_{\eta}=\bar f^i_{\eta}-e_{\eta} |v_{\eta}| \frac{\partial \mc S_{HS}}{\partial \bar f^i_{\eta}},
\end{split}
\end{align}
where $\mc S_{HS}$ is given by Eq.~\eqref{eq:GL_action}, $0<e_\eta<1$ controls the step size and we have multiplied the increment of the field by the absolute value of the eigenvalues $|v_\eta|$ to make $e_\eta$ dimensionless. It is worth noting that  setting $e_\eta = 1$ in these equations   is equivalent to the standard iteration technique  [but for Eq.~\eqref{eq:FFbar before} rather than Eq.~\eqref{eq:Eliashberg}]. 

In the general case the action in Eq.~\eqref{eq:GL_action} is complex. 
However, the equivalence to Eliashberg's equations, Eq.~\eqref{eq:D1 and D2}, implies that the action is real at the saddle-point solution. Without loss of generality we can fix the gauge of the fields such that  $f_{\eta_-} '$  are purely real, $f_{\eta_+} '$  are purely imaginary, and $f_\eta '' = 0$ [as shown in Fig.~\ref{fig:GradientPassing}], which implies that $\D_1$ and $\bar \D_2$ are real and equal to each other. This corresponds to the standard gauge choice which is used in Eliashberg's theory~\cite{margine2013anisotropic}. It should be noted, however, that the fields $f_\eta$ and $\bar f_\eta$ can deviate from this gauge choice during the intermediate steps of the gradient descent method using Eqs.~\eqref{eq:grad_passing_update}.

Let us now demonstrate this procedure on the specific study case, Eq.~\eqref{Eq:V_AM}. In this case the interaction does not depend on momentum and Eqs.~\eqref{eq:grad_passing_update} yield 
\begin{align}\label{eq:gradient passing eqs} 
\begin{split}
           f^{i+1}_{\eta} &=f^{i}_{\eta} -e_{\eta} |v_{\eta}| \left[{\bar{f}^{i}_{\eta}\over |v_{\eta}|}-\zeta_\eta\sum_{\omega} {\frac{\pi \sqrt{TL^3}N_F U^*_{\eta,\w}\bar{\Delta}^{i}_{2}}{\sqrt{\omega^2+\Delta^{i}_1\bar{\Delta}^{i}_2}}} \right] \,,            
           \\           
           \bar{f}^{i+1}_{\eta} &=\bar{f}^{i}_{\eta} -e_{\eta} |v_{\eta}| \left[{f^{i}_{\eta}\over |v_{\eta}|}-\zeta_\eta\sum_{\omega} {\frac{ \pi \sqrt{TL^3}N_FU_{\eta,\w}\Delta^{i}_{1}}{\sqrt{\omega^2+\Delta^{i}_1\bar{\Delta}^{i}_2}}} \right] \,.  
    \end{split}
\end{align}
We find that  the gradient descent method is stable and converges quickly for all values of the repulsion $\mu$ when the step size $e_\eta$ is small enough. 
We demonstrate this in Fig. \ref{fig:Convergent}, where we compare the number of iterations needed to obtain a solution to an accuracy of 1\%  for the two methods, the gradient descent method with $e_\eta = 0.1$ and a straightforward iteration of Eliashberg's equation, as a function of $\mu$ for $\lambda =1$.
In both methods the temperature is  $2\pi T/\w_D =0.03$ and we use a sharp ultraviolet cutoff at $\w_c = 10\w_D$.
We initiate all $f_\eta$ equal to one another and real. We also truncate the number of eigenvalues to $\eta_c = 30$. The  Eliashberg iterative solver is initiated with $\D(\w) = \text{const}$. (See Appendix \ref{app:numerics} for more details.) Figure~\ref{fig:Convergent} shows that when $\mu$ becomes greater than a critical value (in this case, $\mu \approx 0.57$)  the number of iterations needed to obtain a solution by iteration of the Eliashberg equation diverges, while the performance of the gradient descent method is unaffected. For values of $\mu$ greater than the critical value the iterative solution oscillates between negative and positive values and never converges. Interestingly, this break down is not abrupt. As the critical value is approached, the performance of the iterative technique continuously deteriorates. }
 

Let us now discuss the properties of the saddle point that we obtain using the gradient { descent} method. 
In Fig.~\ref{fig:GapWdVsMuGD}(a) we plot the numerical solution of Eqs.~\eqref{eq:gradient passing eqs}, which is expressed as  $\Delta_1(\omega)/\omega_D$ using Eq.~\eqref{eq: D1 and D2 def}, in the space of $\lambda$ and $\mu$. Here we use the  interaction given by Eq.~\eqref{Eq:V_AM} at temperature $2\pi T = 0.03 \omega_D$.  As mentioned above, at the saddle-point solution the action is real and thus 
$\D_1$ is the complex conjugate of $\bar \D_2$. When initiating the search with all $f_\eta$ real and equal we arrive at such a saddle point where  $\D_1$ and $\bar \D_2$ are real and equal, and
\begin{align}
\begin{split}
   \mathrm{Re}[ f_{\eta_-}' ]\ne 0\;\;&\&  \;\; \mrm{Im}[ f_{\eta_-}' ]=0,\\
  \mrm{Re}[ f_{\eta_+}' ]= 0\;\;&\& \;\; \mrm{Im}[ f_{\eta_+}' ]\ne 0,\\
  f_\eta '' &= 0.
  \end{split}
\end{align}
As mentioned above, this is equivalent to the gauge choice in standard Elaishebrg solutions~\cite{margine2013anisotropic}.  Additionally, the odd-frequency modes, $U_{\eta,\w} = - U_{\eta,-\w}$, do not contribute to this solution, so the gap function is symmetric, $\D_i(\w)=\D_i(-\w)$. 
In Fig.~\ref{fig:GapWdVsMuGD}(b) we plot the only repulsive channel $\mrm{Im}[f_{\eta_+}']$ vs. the largest attractive channel $\mrm{Re}[f_{\eta_-}']$, at the saddle point, as $\mu$ is  increased (arrows), for different $\lambda$. This figure visualizes how the location of the  saddle point evolves for different $\mu$.


{ Before concluding this section we comment on the physical consequences of the result we have obtained. Namely, we notice that the solution   exhibits a surprising behavior at large $\lambda$.  $T_c$ remains finite [i.e., higher than the temperature used to generate Fig.~\ref{fig:GapWdVsMuGD}(a)] for arbitrarily large  repulsion $\mu\to \infty$. This behavior was noticed and discussed by the authors of Ref.~\cite{pimenov2022twists} (see Fig. 5 therein). 
In Section~\ref{sec:SE}, we will show that this behavior is an artifact resulting from the omission of the normal-state self-energy corrections.
}


\section{Inclusion of the normal-state self-energy corrections}\label{sec:SE}
The interaction in Eq.~\eqref{Eq:S_I} is non-generic because it only contains scattering in the singlet channel. In order to consider a more generic situation let us use a standard density-density interaction, which has the form  
\begin{widetext}
\begin{align}\label{eq:S_I Full}
\mc S_{\text{int}}={T\over 2L^3}\sum_{\s,\s'}\sum_{\substack{k_1,k_2,\\k_3,k_4}}\hat{V}\left({k_1+k_4\over 2}-{k_2+k_3\over 2}\right) \psi^{\dagger}_{k_1,\sigma} \psi_{k_2,\sigma}\psi^{\dagger}_{k_3,\sigma'}\psi_{k_4,\sigma'} \cdot \delta_{k_1+k_3,k_2+k_4}\,,
\end{align}
{where $\sigma, \sigma' = \uparrow, \downarrow$ denote electron's spin.} Clearly, there are contributions to this interaction that do not appear in Eq.~\eqref{Eq:S_I}. 
These contributions are detrimental to spin-singlet superconductivity and must therefore be taken into account. 
The authors of 
Ref.~\cite{Protter2021Functional} showed that these terms modify the action and its saddle-point equations to include the normal-state self-energy corrections, as in Eq.~\eqref{eq:Eliashberg}. This is done by performing the HS transformation with the additional decoupling field in the particle-hole channel.  

To see how this works, we divide Eq.~\eqref{eq:S_I Full} into two contributions, $\mc S_{\text{int}} = \mc S_I+\mc S'_I$, with $\sigma' = -\sigma$ and $\sigma'=\sigma$, respectively. When time-reversal and inversion symmetries are present the former contribution assumes the form of Eq.~\eqref{Eq:S_I}, 
while the latter is given by
\begin{align}\label{eq:S_I'}
    \mc S_{I}'=-\frac{T}{2L^3}\sum_{Q,k,p,\sigma} \psi^{\dagger}_{k-\frac{Q}2,\sigma}\psi_{k+\frac{Q}2,\sigma} \hat{V}_{k,p} \psi^{\dagger}_{p+\frac{Q}2,\sigma}\psi_{p-\frac{Q}2,\sigma}\,.
\end{align}
Note that in each one of these contributions we breakdown the delta-function, implementing momentum conservation in different manners. Namely, in Eq.~\eqref{Eq:S_I} we use $k_1+k_3=k_2 + k_4 = Q$,
while in Eq.~\eqref{eq:S_I'} we use $k_4 - k_1 = k_3 - k_2 = Q$.
Also notice the minus sign on the RHS of Eq.~\eqref{eq:S_I'}, which comes from {anticommuting the} Grassmann fields. 

Analogously to Eq.~\eqref{Eq:S_I}, we rewrite interaction \eqref{eq:S_I'} in terms of fermionic bilinears { in the particle-hole channel}, {  $\Gamma_{k,\sigma}(Q)=\psi^{\dagger}_{k+\frac{Q}2,\sigma}\psi_{k-\frac{Q}2,\sigma}$}, which gives
\begin{align}
    \mc S_{\text{int}}={T\over L^3}\left[\sum_{Q,k,p} \Lambda^\dagger_k(Q)\hat{V}_{k,p} \Lambda_p(Q)-\frac{1}{2}{ \sum_\sigma}\sum_{{ Q,} k,p} \Gamma^\dagger_{k,\sigma}(Q)\hat{V}_{k,p}\Gamma_{p,\sigma}(Q)\right]\,.
\end{align}
\end{widetext} 
Then, we use  Eq.~\eqref{eq:Intkp} to transform to the diagonal basis 
\begin{align}
    \mc S_{\text{int}}=\sum_{\eta,Q} v_{\eta}\left[\varphi^\dagger_{\eta}(Q)  \varphi_{\eta}(Q)-\frac{1}{2}{\sum_\sigma}\gamma^\dagger_{\eta,\sigma}(Q)\gamma_{\eta,\sigma}(Q)\right]\,,
\end{align}
where $\varphi_{\eta}(Q)=\sqrt{T/L^3}\sum_{k}\Lambda_{k}(Q)U_{\eta,k}$ as before 
and $\gamma_{\eta,\sigma}(Q)=\sqrt{T/L^3}\sum_{k}\Gamma_{k,\sigma}(Q)U_{\eta,k}$. We also note that when the eigenvectors of the interaction are real, i.e., ${U}^*_{\eta,k} = {U}_{\eta,k}$, then $\gamma^\dagger_{\eta,\sigma}(Q) = \gamma_{\eta,\sigma}(-Q)$, implying that $\gamma_{\eta,\sigma}$ is real.

Next, we perform the HS transformation. 
{The transformation in the particle-particle channel is described in Section~\ref{Sec:HS transformation}, where the fields $\varphi_\eta$ are coupled to the { complex} bosonic auxiliary fields $f_{\eta}$ with the coupling $\z_\eta$. Since the bilinears in the particle-hole channel $\gamma_{\eta,\sigma}$ are real, they are coupled to the \textbf{real} bosonic field satisfying $g^*_{\eta,\sigma}(Q) = g_{\eta,\sigma}(-Q)$, with the coupling $i\z_\eta$.} The resulting action is given by 
{
\begin{widetext}
\begin{equation}
    \mc S_{HS}=S_0+\sum_{\eta,Q} \left\{\frac{|f_{\eta}(Q)|^2}{|v_\eta|}  -\zeta_{\eta}\left[f^*_{\eta}(Q)\varphi_{\eta}(Q)+f_{\eta}(Q)\varphi^\dagger_{\eta}(Q)\right]+ \sum_{\sigma}\frac{|g_{\eta,\sigma}(Q)|^2}{2|v_\eta|} 
   +i\z_{\eta}g_{\eta,\sigma}(Q)\gamma_{\eta,\sigma}(Q)\right\},
\end{equation}
where $S_0$ denotes the free-fermionic part.
Finally, integrating out the fermions we obtain the bosonic action 
\begin{align}
\mc S_{HS}=\sum_{\eta,Q} \frac{|f_{\eta}(Q)|^2}{|v_\eta|} +\frac{|g_{\eta,\uparrow}(Q)|^2 + |g_{\eta,\downarrow}(Q)|^2}{2|v_\eta|} -\tr{\ln{\mc G_{k}^{-1}(Q)}}\,,
\end{align}
\end{widetext}
where the Green's function in Nambu space is given by
\begin{align}
    \mathcal G_{k}^{-1}(Q)= 
    \begin{pmatrix}
G_\uparrow^{-1}(k,Q)  & -\Delta_{1,k}(Q) \\
-\bar\Delta_{2,k+Q}(-Q) & -G_\downarrow^{-1}(-k-Q,Q)
\end{pmatrix}.
\end{align}
Here we defined
$G^{-1}_{\sigma}(k,Q)=G^{-1}_0(k) \delta_{Q,0}+\Sigma_{\sigma,k}(Q)$, with 
\begin{align}
\Sigma_{\sigma,k}(Q)=i\sqrt{T\over L^3}\sum_{\eta}\zeta_\eta U_{\eta,k+\frac{Q}2}g_{\eta,\sigma}(Q) .
\end{align}}

Let us now explore the $Q=0$ saddle point of the action with respect to the fields $g_{\eta,\sigma}$, which in this case become purely real due to the identity $g_{\eta,\sigma}(0) = g^*_{\eta,\sigma}(0)$. As an example, taking the derivative with respect to  $g_{\eta,\uparrow}$ gives
\begin{align}
    \frac{\partial S_{HS}}{\partial g_{\eta,\uparrow}}=\frac{g_{\eta,\uparrow}}{|v_{\eta}|}-\sqrt{T\over L^3}\,\tr{\left[\mc{G}_{k}(0)
    \begin{pmatrix}
        i\z_{\eta} U_{\eta,k} & 0\\
        0 & 0
    \end{pmatrix}\right]}=0
\,.\end{align}
Therefore, the saddle point equations for $g_{\eta,\sigma}$ are given by
\begin{widetext}
\begin{align}
\begin{split}
      \frac{\partial S_{HS}}{\partial g_{\eta,\uparrow}}=\frac{g_{\eta,\uparrow}}{|v_{\eta}|}+i\z_{\eta}
     \sqrt{\frac{T}{L^3} }\sum_{k}\frac{(-i\omega+\Sigma_k)U_{\eta,k}-(\xi_k+\chi_k)U_{\eta,k}}{(\xi_k+\chi_k)^2-(-i\omega+\Sigma_k)^2+\Delta_{1,k}\bar\Delta_{2,k}}=0,
     \\
     \frac{\partial S_{HS}}{\partial g_{\eta,\downarrow}}=\frac{g_{\eta,\downarrow}}{|v_{\eta}|}+i\z_{\eta}
     \sqrt{\frac{T}{L^3} }\sum_{k}\frac{(i\omega-\Sigma_k)U_{\eta,-k}-(\xi_k+\chi_k)U_{\eta,-k}}{(\xi_k+\chi_k)^2-(-i\omega+\Sigma_k)^2+\Delta_{1,k}\bar\Delta_{2,k}}=0\,,
\end{split}
\end{align}
where we used the definitions
\begin{align*}
\Sigma_k\equiv{\Sigma_{\uparrow,k}(0)-\Sigma_{\downarrow,-k}(0)\over 2};\;\;\;\;\;\;\;\;
\chi_k\equiv{\Sigma_{\uparrow,k}(0)+\Sigma_{\downarrow,-k}(0)\over 2}\,.
\end{align*}
These notations coincide with the ones commonly used in standard Eliashberg theory~\cite{marsiglio2020eliashberg}.  Then, using these equations, we can derive  Eliashberg's equations for the normal (diagonal) part of the self-energy, analogously to how it was done in Section~\ref{Sec:Eleq}
\begin{align}
\Sigma_p &= \frac{T}{L^3}\sum_{k}\frac{V_{p,k}(i\omega-\Sigma_k)}{(\xi_k+\chi_k)^2-(-i\omega+\Sigma_k)^2+\Delta_{1,k}\bar\Delta_{2,k}}, \nonumber \\
\chi_p &= \frac{T}{L^3}\sum_{k}\frac{V_{p,k}(\xi_k + \chi_k)}{(\xi_k+\chi_k)^2-(-i\omega+\Sigma_k)^2+\Delta_{1,k}\bar\Delta_{2,k}}.
\end{align}

Now let us focus on the specific example of Eq.~\eqref{Eq:V_AM}. Following standard approximations used in Eliashberg theory~\cite{margine2013anisotropic}, we neglect the dispersion renormalization $\chi_k$, which is typically justified in the limit where the Fermi energy is much larger than the Debye frequency. Moreover, in the case of the momentum-independent interaction as in Eq.~\eqref{Eq:V_AM}, $\Sigma_k$ and $\Delta_{i,k}$ become functions of frequency only, so one can integrate over momentum explicitly to obtain:
\begin{align}\label{eq:gup=gdw}
\begin{split}
     \frac{g_{\eta,\uparrow}}{\sqrt{L^3T}}=-i  {\pi  N_F|v_{\eta}|\z_{\eta}}\sum_{\omega}\frac{(-i\omega+\Sigma_\omega)U_{\eta,\omega}}{\sqrt{\Delta_{1,\w}\bar\Delta_{2,\w}-(-i\omega+\Sigma_\omega)^2}}, \\
     \frac{g_{\eta,\downarrow}}{\sqrt{L^3T}}=i  {\pi  N_F|v_{\eta}|\z_{\eta}}\sum_{\omega}\frac{(-i\omega+\Sigma_\omega)U_{\eta,-\omega}}{\sqrt{\Delta_{1,\w}\bar\Delta_{2,\w}-(-i\omega+\Sigma_\omega)^2}} \,.
     \end{split}
\end{align}

Time-reversal symmetry, which is assumed to be present in our system, implies further that $g_{\eta,\uparrow} = g_{\eta,\downarrow}$, so only odd-frequency modes, $U_{\eta,\w} = - U_{\eta,-\w}$, contribute to $g_{\eta,\s}$. It is equivalent to the statement that the normal part of the self-energy is odd under frequency,  $\Sigma_\w = - \Sigma_{-\w}$.

The derivatives with respect to $f_\eta$ and $\bar f_\eta$  give the same equations as before, Eqs.~\eqref{eq:FFbar before}, with the standard modifications to the Green's function,  $i\omega \to i\omega - \Sigma_k$ and $\xi_k \to \xi_k + \chi_k$. Once again, under the assumptions made above, we integrate over momenta and obtain 
\begin{align}\label{eq:fandfbar with SE}
\begin{split}
           {f_\eta\over \sqrt{L^3T}} &=\pi N_F  |v_\eta|\zeta_{\eta}\sum_{\omega} {\frac{\Delta_{1,\omega} U_{\eta,\omega}}{\sqrt{\Delta_{1,\w}\bar \Delta_{2,\w}-(-i\omega+\Sigma_\omega)^2}}},\\                
           {\bar{f}_\eta\over \sqrt{L^3T} }&=\pi  N_F|v_\eta| \zeta_{\eta}\sum_{\omega} {\frac{\bar\Delta_{2,\omega}U^*_{\eta,\omega}}{\sqrt{\Delta_{1,\w}\bar\Delta_{2,\w}-(-i\omega+\Sigma_\omega)^2}}}\,. \\
\end{split}
\end{align}
\end{widetext}
Together these equations define the saddle point of the action including the normal self-energy corrections.

The numerical solution of the saddle-point equations~\eqref{eq:gup=gdw} and \eqref{eq:fandfbar with SE} is obtained using the gradient { descent} method described in Section~\ref{sec:GD}. The result is presented in Figs.~\ref{fig:GapWdVsMuGD}(c) and (d). In panel (c) we plot the order parameter $\D_1(0)$, defined in Eq.~\eqref{eq: D1 and D2 def}, as a function of $\l$ and $\mu$ from Eq.~\eqref{Eq:V_AM}. Comparing with panel (a), we find that the inclusion of self-energy corrections is crucial, especially in the limit of large $\l$. In particular, { it seems} to cure the unphysical behavior in this regime by diminishing $T_c$ to zero at a sufficiently large $\mu$ for all $\l$. In panel (d) we plot the solution for the order parameters  $\Sigma(\w)$ (top panel) and $\D_1(\w)$ (bottom panel) for $\l = 1.3$ and $\mu=0.5$ (marked by the black asterisk in panel (c)).

\section{Fluctuations around the saddle point and derivation of a GL theory}\label{sec:GL theory}
We now consider the fluctuations around the saddle point described in the previous sections, { following the line of Ref.~\cite{Protter2021Functional}}. To capture their contribution  one must parametrize the field's fluctuation  to be  along the direction of  steepest descent in the complex plane~\cite{bender1999advanced}. Below $T_c$ the saddle point is generally located somewhere in the complex plane, which requires additional care. However, in this  work we will  mostly  consider the  case where $T$ is close to, but higher than $T_c$. The saddle-point solution is then trivially zero and is located on the real axis. Nonetheless, the direction of steepest descent  may still extend into the complex plane.

In the most generic situation, we expand the fields $f_\eta$ relative to their saddle-point solution given by Eq.~\eqref{eq:gapFFbar},\footnote{We neglect the normal-state self-energy corrections in this section. However, such corrections can most definitely become important. We leave their inclusion to future work. } which we denote henceforth as $f_\eta^{(0)}$:
\begin{align}\label{eq:f expnasion}
    f_{\eta}(Q)&=f^{(0)}_{\eta}+a_{\eta}(Q)+ib_\eta(Q)\,.
\end{align}
Here $a_{\eta}$ and $b_{\eta}$ are complex fluctuations of the fields $f'_{\eta}$ and $f''_{\eta}$ in Eqs.~\eqref{eq:f' and f'' to f and fbar} and \eqref{eq:f* to bar f}, respectively. 
The corresponding order parameters in momentum-frequency space, $\Delta_{1}$ and $\bar\D_2$, are also  written relative to their saddle-point values 
\begin{align} \label{Eq:deltaab}
    \Delta_{1,k}(Q)-&\Delta^{(0)}_{1,k}=\delta_{1,k}(Q)\\&
    =\sqrt{T\over L^3}\sum_{\eta}\zeta_{\eta}U^*_{\eta,k} [a_{\eta}(Q)+ib_{\eta}(Q)]\,,\nonumber \\
    \bar \Delta_{2,k}(Q)-&\bar \Delta^{(0)}_{2,k}=\bar \delta_{2,k}(Q) \nonumber \\&=\sqrt{T\over L^3}\sum_{\eta}\zeta_{\eta}U_{\eta,k}[a_{\eta}(Q)-ib_{\eta}(Q)]\,.\nonumber
\end{align}

To obtain the GL theory above $T_c$ we set $f_\eta ^{(0)}=0$ in Eq.~\eqref{eq:f expnasion}. We then expand the action \eqref{eq:GL_action} to quadratic order in fluctuations  $a_{\eta}$ and $b_{\eta}$, which yields 
\begin{align}\label{eq:S2_before symmetrization}
    \mc S_{GL}^{(2)}= \sum_{Q, \eta, \eta'}\vec{\alpha}_{\eta }^T(Q)
   \hat M_{\eta ,\eta'}(Q)
    \vec{\alpha}_{\eta'}(Q)\,,
\end{align}
where 
\begin{align}\label{eq:M}
\hat M_{\eta ,\eta'}(Q) = \hat{V}_{\eta ,\eta'}^{-1}-\hat{S}_{\eta ,\eta'}(Q) 
\end{align}
is the fluctuation matrix and $\vec{\alpha}_\eta^T(Q) = [a_\eta(Q),b_\eta(Q)]$ is the the vector of fluctuation fields.
The two matrices composing the fluctuation matrix in Eq.~\eqref{eq:M} are given by
\[\hat{V}^{-1}_{\eta,\eta'} = |v_\eta|^{-1}\delta_{\eta,\eta'}\begin{pmatrix}
       1& 0 \\
       0  & 1 
         \end{pmatrix}\] 
and
\begin{align}\label{eq:S_matrix_app}
  \hat{S}_{\eta \eta'}(Q)=B_{\eta,\eta'}(Q)
    \begin{pmatrix}
       1& -i \\
       i  & 1 
         \end{pmatrix}\,,
\end{align}
where
\[
B_{\eta,\eta'}(Q)=\zeta_{\eta}\zeta_{\eta'}\frac{T}{L^3}\sum_{k}U^*_{\eta,k}G_0(k)G_0(-k-Q)U_{\eta',k} \,,\]
while $G_0(k)$ is defined below Eq.~\eqref{Eq:SHS}. The resulting fluctuation matrix $\hat M(Q)$ is a $2N_\eta \times 2N_\eta$  matrix, which becomes non-Hermitian in the presence of repulsion, and $N_\eta$ is the number of eigenchannels in Eq.~\eqref{eq:eigs_decomp}. However, only the symmetric part of this matrix contributes to the action, as can be seen from Eq.~\eqref{eq:S2_before symmetrization} (for more details see Appendix \ref{app:FulctMatrix}). Thus, we can replace the  matrix $\hat M$ with its symmetric part $\hat M_s =(\hat M+\hat M^T)/2 $. Consequently,  Eq.~\eqref{eq:S2_before symmetrization} assumes the form 
\begin{align}\label{eq:S^2_GL}
    &\mc S_{GL}^{(2)}= \sum_{Q}\vec{\alpha}^T(Q)
    \hat M_s(Q)
    \vec{\alpha}(Q)\,.
\end{align}

The Autonne-Takagi (AT) factorization~\cite{autonne1915matrices,takagi1924algebraic} ensures that when the fluctuation matrix can be diagonalized, it can be done using a unitary matrix $\mathcal W$, such that the diagonal elements are  real non-negative numbers.
Namely, because the matrix is symmetric, it is diagonalized by an orthogonal matrix $W$, $\hat M_s=W^T\, \hat \varepsilon\, W$, where  
\begin{align}\label{eq:eps_hat}
\hat \varepsilon = \mrm{diag}(\varepsilon_1 e^{i\phi_1},\varepsilon_2e^{i\phi_2},\ldots,\ve_{2N_\eta}e^{i\phi_{2N_\eta}})
\end{align} 
 is the diagonal eigenvalue matrix and $\bs X = W\bs \a$ is the vector of  fluctuation eigenmodes.  
We can then multiply this vector by a diagonal matrix of phases $P$ that counters the phases $\phi_j$ of the eigenvalues, 
\[\bs{X} = P^{-1} \tilde{\bs X} \;\;\to\;\;X_j = e^{-{i\over 2}{\phi_j }}\tilde X_{j}\,,\] 
such that $\mathcal W = PW$ is the AT unitary transformation, while the diagonalized fluctuation matrix consists of the real absolute values $\ve_j \ge 0$. Generic values of the phases $\phi_j$ merely define the steepest descent direction for the corresponding fields in the complex plane and should not be associated with  mode dissipation.  

However, because the matrix $\hat M_s$ is non-Hermitian, there might be points in the parameter space where it becomes defective in the sense that it cannot be diagonalized. These points are known as {\textit{exceptional points}}~\cite{kato1966analytic,doi:10.1126/science.aar7709}. In what follows we will see that such exceptional points appear in the field theoretic description of superconductivity when  repulsion  is present. These points can be tuned by different parameters such as temperature or the center of mass momentum $Q$.

Finally, we note that the eigenvalues of the matrix $\hat M_s$ are doubly degenerate, which is important to ensure a gauge invariant bosonic theory. Let us focus on the eigenvalue with the smallest absolute value $\ve_m$ and the two corresponding  eigenmodes  $\tilde X_1$ and $\tilde X_2$. 
This pair will form the real and imaginary parts of the Ginzburg-Landau order parameter. Namely 
\[ \Psi(Q) = \tilde X_1(Q) + i\tilde X_2(Q)\,, \]
where $\Psi(Q)$ is proportional to the conventional Ginzburg-Landau field. Neglecting quantum fluctuations, we then perform a spatial gradient expansion and obtain the GL theory 
\begin{align}\label{eq:free energy}
    \mc S_{GL}^{(2)}=\int d\bs x\left[ \varepsilon_{m}(0)|\Psi|^2 + {\varepsilon_{m}''(0)\over 2}{|(\nabla-2ie\bs A)\Psi|^2}+\ldots \right]. 
\end{align}
As mentioned above, the values $\ve_j$ are positive by construction. The superconducting transition point, which coincides with Eliashberg's theory, is obtained when $\ve_m(0)=0$. Below this temperature the analysis we have performed here is no longer valid and an expansion around the new saddle point with $f_\eta^{(0)}\ne0$ is required.  

Equation~\eqref{eq:free energy} describes the long wavelength properties of the superconductor above $T_c$, and in particular how they depend on the microscopic parameters of the pairing interaction, Eq.~\eqref{Eq:V_AM}. To demonstrate this { with the implication to}  experimental observables, we will focus specifically on the upper critical field  $H_{c2}$ (close to $T_c$)\footnote{In the standard analysis of the Ginzburg-Landau theory the value of $H_{c2}$ close to $T_c$ is obtained by taking the mass term ($\ve_m(0)$ in our case) to be infinitesimally small and \textit{negative}. The GL equation then become equivalent to that of a quantum harmonic oscillator where $-\ve_m(0)$ plays the role of the positive ground state energy.  However, here $\ve_m(0)$ is positive by construction, since the system is above $T_c$. Thus, to extract $H_{c2}$ we make the assumption that the slope of the approach of the eigenvalue  $\ve_m(0)= r (1-T_c/T)$ is the same on both sides of the transition and therefore the slope  { above $T_c$} can indicate the  asymptotic behavior { below $T_c$}. Whether this is true remains to be verified in a future publication where we will  {derive} the GL theory on the {superconducting} side of the transition.} , which is given by~\cite{tinkham2004introduction}
\begin{align}\label{eq:Hc2}
    H_{c2} = {\Phi_0 \over 2\pi\xi_{GL}^2}(1-T/T_c), 
\end{align}
where $\Phi_0 = h/2e$ is the flux quantum and asymptotic behavior of the ratio 
\[\xi_{GL}^2 =  {\varepsilon_m''(0)\over 2\varepsilon_m(0)}(1-T/T_c),\;\; T\to T_c\]  
is the GL coherence length. 

Indeed, within our quadratic approximations, the inclusion of repulsion in the pairing interaction leads to distinctive features in the upper critical field, $H_{c2}$. In particular, the eigenvalue controlling the superconducting transition, $\ve_m$ in Eq.~\eqref{eq:free energy}, exhibits an exceptional point that is tuned by the repulsion strength and causes $H_{c2}$ to peak at a critical value. In what follows, we will demonstrate this feature on two types of pairing interactions and over a wide parameter range showing that it is a robust feature of pairing interactions with repulsion.



\begin{figure}
    \centering
    \includegraphics[width=\linewidth]{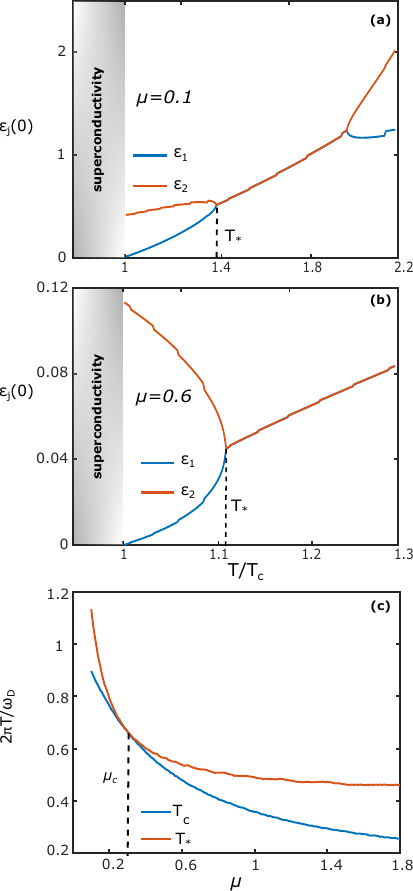}
    \caption{The absolute values controlling the fluctuations of the Ginzburg-Landau theory for the toy model described by Eq.~\eqref{eq:ToyModel} (see also Appendix~\ref{app:ToyModel}).  (a), (b) The two absolute values in Eq.~\eqref{eq:eps_hat}, $\ve_{1,2}(0)$, at $Q=0$ as a function of $T/T_c$ for $\mu$ below ($\mu = 0.1$) and above ($\mu = 0.6$) the critical value $\mu_c\approx 0.3$, respectively. Both $T_c$ (SC transition temperature) and $T_*$ (exceptional point temperature) are marked on the plots. The ``staircase'' structure is unphysical and appears due to hard frequency cutoff (see Appendix \ref{app:cutoff}). (c) The values of $T_c$ and $T_*$ as a function of $\mu$. Note that $\mu_c$, the value of repulsion where $T_c=T_*$, is marked on the plot. Also note that for numerical convenience we have used a large coupling strength $\lambda=1.6$ in these plots. }
    \label{fig:eigs_fluctuation_matrix}
\end{figure}

\begin{figure*}
    \centering
    \includegraphics[width=\textwidth]{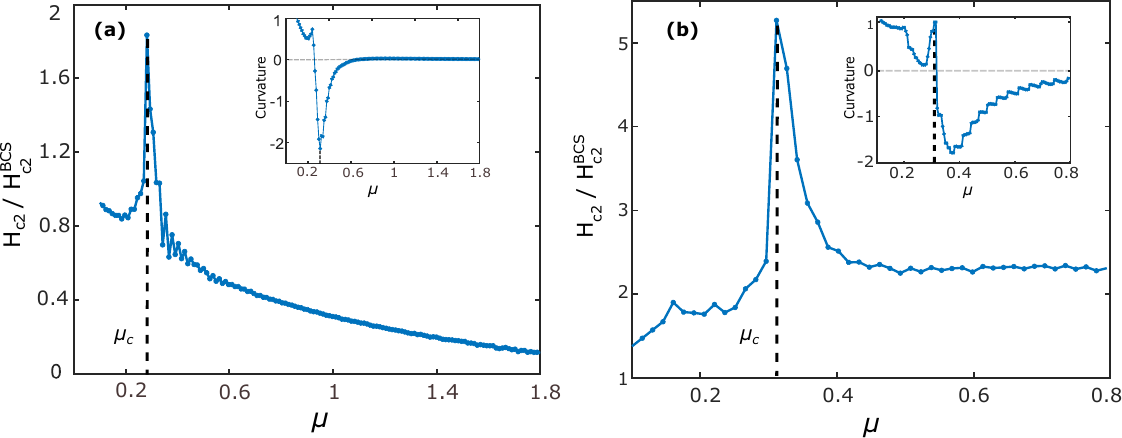}
    \caption{The ratio between Eq.~\eqref{eq:Hc2} and Eq.~\eqref{eq:Hc2^BCS}  as a function of $\mu$ for (a) the toy model, Eq.~\eqref{eq:ToyModel}, and (b) Anderson-Morel interaction, Eq.~\eqref{Eq:V_AM}. This quantity is essentially ratio between the linear slope of $H_{c2}$ close to $T_c$ relative to the expectation from BCS theory. Note that for each value of $\mu$ Eq.~\eqref{eq:Hc2^BCS} is taken at $T_c(\mu)$, which is a monotonically decreasing function of $\mu$.   
    The curvature (i.e., $T^2_c\frac{d^2H_{c2}}{dT^2}\big|_{T_c}$) is plotted in the inset. The $H_{c2}$ in these plots has been calculated for $\lambda=1.6$ and $\omega_c=30 \omega_D$. The large value of $\lambda$ is used to minimize cutoff effects resulting from the discrete Matsubara sum. We have verified that the results do not change qualitatively for smaller coupling. We note that at $\lambda = 1.6$ the gap and $T_c$  depend very weakly on $\mu$ when the normal-state self-energy corrections are not taken into account, as shown in Fig.~\ref{fig:GapWdVsMuGD}. This is the reason why $H_{c2}$ saturates to a finite value at large $\mu$.  } 
    \label{fig:Hc2}
\end{figure*}

\subsection{Results for a simplified toy model with two eigenvalues }
{We first demonstrate the $H_{c2}$ calculation from the GL theory given by Eq.~\eqref{eq:free energy} on the simple example of a toy model interaction:}
\begin{align}
\label{eq:ToyModel}
\hat V_{\w,\w'} = {\lambda \over N_F}\left[ \mu - {1\over 1+ (\w/\w_D)^2}{1\over 1+ (\w'/\w_D)^2 } \right]\,.
\end{align} 
This interaction is designed to be similar to Eq.~\eqref{Eq:V_AM}, and has the advantage of having only two non-vanishing eigenvalues $v_\eta$, one repulsive and one attractive. However, it is clearly not time-translationally invariant.

The interaction in Eq.~\eqref{eq:ToyModel}  does not depend on momentum. Furthermore, we will only be interested in the static GL free energy. As a consequence we can compute the fluctuation matrix analytically  (see Appendix~\ref{app:PairingSucept}). 

In Figs.~\ref{fig:eigs_fluctuation_matrix}(a) and (b) we plot the two absolute values $\varepsilon_j(0)$ extracted after diagonalization of the fluctuation matrix $\hat M_s(Q)$, as a function of temperature for two different values of the repulsion, $\mu=0.1$  and $\mu=0.6$, respectively and $\lambda = 1.6$.\footnote{This relatively large value of the coupling was chosen to minimize cutoff effects. The results presented in this section appear also in the weak coupling limit.}   
The spectrum is at least doubly degenerate due to gauge invariance, so the two smaller absolute values are labeled $\ve_1$ and the two larger ones are labeled by $\ve_2$. The former corresponds to the values that vanish at $T=T_c$, i.e., $\ve_m$ in Eq.~\eqref{eq:free energy}.
We also note that there is a temperature $T_*$, where an exceptional point occurs. This point is manifested by the coalescence of the eigenvalues (also marked by a dashed line).\footnote{The jumps in the curve are a numerical artifact stemming from the cutoff $\w_c$. We have smoothed this effect by taking a large cutoff and by softening the cutoff (see Appendix ~\ref{app:cutoff}).} 
In Fig.~\ref{fig:eigs_fluctuation_matrix} (c)  we plot $T_c$ and $T_*$ as a function of the repulsion strength $\mu$. Note that $T_*\ge T_c$ for all $\mu$. Interestingly however, there is a critical value of the repulsion $\mu_c\approx 0.3$ where  the two temperatures touch. At this critical value the matrix is defective at $T_c$.

In Fig.~\ref{fig:Hc2}(a) we plot the upper critical field, Eq.~\eqref{eq:Hc2}, normalized by Gor'kov's expression for a BCS superconductor~\cite{gor1959microscopic}
\be\label{eq:Hc2^BCS}
H_{c2}^{BCS} \approx {24 \pi  T_c^2 \Phi_0^2 \over 7 \zeta(3)v_F^2}\left(1-{T\over T_c}\right), \;\; T\to T_c.
\ee
Note that $T_c$ appearing here is a function of $\mu$, as shown in Fig.~\ref{fig:eigs_fluctuation_matrix}(c). 
As can be seen,  $H_{c2}$ is approaching is Gor'kov's prediction in the limit of small $\mu$~\footnote{However, we do not expect the result to become exactly identical in the limit $\mu\to0$. This is because the result in Eq.~\eqref{eq:Hc2^BCS} is obtained with a contact interaction with just one eigenvector $U_{\eta,k}=\text{const}$ with nonzero eigenvalue, while both models considered in this paper have nontrivial frequency dependence.}.
However, upon increasing $\mu$,
$H_{c2}$ shows a non-monotonic behavior, peaking around the critical value $\mu_c$ before diminishing significantly compared to the expectation from BCS theory, Eq.~\eqref{eq:Hc2^BCS}.  The origin of the peak is the rapid variation of the numbers $\ve_j(Q)$ with temperature near the exceptional point. 
That is, they  depend strongly on temperature close to the transition when $T_c$ and $T_*$ are close. Interestingly, we find that both the mass term $\varepsilon_m(0)$ and the second derivative $\varepsilon_m''(0)$ develop singular behavior near $\mu_c$ (i.e., they are nonanalytic in the variable $1-T/T_c$). However, the ratio in Eq.~\eqref{eq:Hc2} remains linear, leading to a finite ratio of  Eq.~\eqref{eq:Hc2} with  Eq.~\eqref{eq:Hc2^BCS} in the limit $T\to T_c$. 
The inset of Fig.~\ref{fig:Hc2} (a) displays the curvature of $H_{c2}$ as it approaches zero near $T_c$. Here it should be noted that we simply compute the second derivative of $\xi_{GL}^{-2}$ with respect to temperature. Whether this gives the correct asymptotic series on the ordered side of the transition remains to be checked. We find that $\mu_c$ also marks a transition between  positive and negative curvature of the asymptotic curve.

Above we predicted that $H_{c2}$ will sharply peak at a critical value of the repulsion (assuming this parameter can be tuned experimentally). 
However, it should be noted that the approximations used to obtain $H_{c2}$  can breakdown in the vicinity of $\mu_c$ due to a number of reasons. The singular temperature dependence of both the homogeneous mass term and the second derivative implies that higher order terms may also become singular (e.g., the quartic term in the fields or higher derivatives in the expansion). These higher order terms need to be carefully compared with the second order terms. Moreover,  the two absolute values controlling the fluctuation matrix $\ve_{1,2}$ correspond to two distinct modes in the GL theory. 
These are degenerate at $T_c$ when $\mu=\mu_c$. Thus a multi-mode GL theory must be employed. However, it should also be noted that at $\mu_c$ the fluctuation matrix is defective,  raising a question regarding the nature of such  modes. We conclude that the inclusion of these effects in our theory may  modify the result for the upper critical field compared to the quadratic approximation presented in Fig.~\ref{fig:Hc2}. For example, they may remove the sharp peak at $\mu_c$. We leave such an extensive investigation to a future publication.

\subsection{Results for the Morel-Anderson interaction }
After gaining intuition for the influence of repulsion on the upper critical field $H_{c2}$ using the toy model, Eq.~\eqref{eq:ToyModel}, we now go back to the full Morel-Anderson interaction in Eq.~\eqref{Eq:V_AM}. 
In Fig.~\ref{fig:Hc2} (b) we plot the ratio between the asymptotic expressions in Eq.~\eqref{eq:Hc2} and Eq.~\eqref{eq:Hc2^BCS} as a function of $\mu$ for $\lambda = 1.6$ and neglecting normal-state self-energy corrections. 
As for the toy model, the value of $H_{c2}$ converges to Eq.~\eqref{eq:Hc2^BCS} in the limit $\mu\to 0$.
However, we find that the enhancement of $H_{c2}$ near $\mu_c$ is much more prominent and occurs at a similar value of the repulsion (but not the same). Moreover, in this model $H_{c2}$ does not decrease in the limit $\mu \to \infty$, but seems to saturate at a value that is roughly twice the prediction of a BCS theory. The inset shows that  the curvature of the asymptotic expression for $H_{c2}$ near $T_c$ also changes sign at $\mu=\mu_c$.

\begin{figure}
    \centering
    \includegraphics[width=1\linewidth]{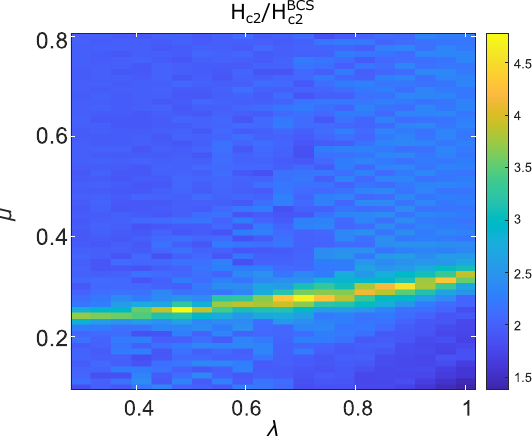}
    \caption{The ratio between Eq.~\eqref{eq:Hc2} and Eq.~\eqref{eq:Hc2^BCS}  as a function of $\mu$ and $\lambda$ for Anderson-Morel interaction, Eq.~\eqref{Eq:V_AM}.  The enhancement of $H_{c2}$ near $\mu_c$ is robust and remains for a wide range of the value of $\lambda$. The ``noisy'' feature in the heat-map is a numerical artifact and the consequence of the frequency cutoff (as explained in Appendix~\ref{app:cutoff}).  } 
    \label{fig:Hc2HeatMap}
\end{figure}

As mentioned, here we have used a large coupling $\lambda = 1.6$ to minimize cutoff effects. In this limit, however, 
the gap and $T_c$ depend very weakly on $\mu$ as long as the normal-state self-energy corrections are not taken into account, as shown in Fig.~\ref{fig:GapWdVsMuGD} (a). This is the reason that $H_{c2}$ saturates to a finite value at large $\mu$.
To explore a larger range of the coupling $\lambda$, in Fig. \ref{fig:Hc2HeatMap} we plot the ratio $H_{c2}/H^{BCS}_{c2}$ on a color map   as a function of both $\lambda$ and $\mu$. The existence of a critical $\mu_c$ seems to be a universal feature for all $\lambda$. 

Thus, the behavior in the case of Eq.~\eqref{Eq:V_AM} is quantitatively different from what we found for Eq.~\eqref{eq:ToyModel}, but qualitatively similar. Namely, in both cases there exists a temperature $T_*$ where the  eigenvalues with smallest absolute value incur an exceptional point, which bounds $T_c$ from above,  $T_* \ge T_c$. Moreover, in both models there is a critical value $\mu_c$, where the two temperatures touch but do not cross, leading to a peak in $H_{c2}$. These results suggest that the existence of a critical repulsion strength $\mu_c$ is possibly a universal feature of superconductors with repulsion.

\section{Conclusions and Discussion}
We developed a field theoretic description for superconductors which include repulsive interactions  using the Hubbard-Stratonovich transformation. We first decomposed the interaction into eigenchannels. Then we performed the Hubbard-Stratonovich transformation such that repulsive channels were coupled via an imaginary coupling and attractive ones via a real coupling. The resulting action was found to have a saddle point that can be shifted outside the original field-integration line into the complex plane. The saddle point was shown to coincide with Eliashberg's theory and captures the physics of fluctuations around this  solution. 


To numerically obtain the  saddle-point solution we used the gradient descent method, which allows us to update the gap in small increments in the complex plane. This method outperforms a straightforward iteration of Eliashberg's equations when strong repulsion is present. We also incorporated the normal-state self-energy corrections, which hold a crucial role in this limit. After obtaining  the saddle-point solution and understanding its properties, we  proceeded to discuss  fluctuations of the order parameter around this solution. We demonstrated how to derive a theory capturing such fluctuations for the temperature range  above and close to $T_c$ (the Ginzburg-Landau theory). The matrix controlling the Gaussian fluctuations about the saddle point was found to be non-Hermitian due to the presence of repulsive interaction, and the directions of fluctuations in the complex plane were chosen according to the steepest descent method. We applied this theory to calculate the dependence of the upper critical field on the repulsion strength close to $T_c$ for two types of pairing interactions. 

The first type was a toy model interaction that has only two non-zero eigenvalues, one repulsive and  one attractive. The second example was the Morel-Anderson interaction given by Eq.~\eqref{Eq:V_AM}. In both cases we found that the fluctuation matrix has a temperature $T_*$, where it has an exceptional point and hence cannot be diagonalized. This temperature is found to be always greater or equal to $T_c$. Interestingly, in both models there exists a critical value of the repulsion $\mu_c$  such that these two temperatures coalesce. 
The linear slope of $H_{c2}$ as a function of temperature close to $T_c$ was computed in both cases. Within the quadratic approximation, $H_{c2}$ was found to peak at the critical value of repulsion $\mu_c$ due to the existence of the exceptional point. However, we have also cautioned that our approximations can  breakdown near $\mu_c$ due to a number of reasons. Consequently, analysis that goes beyond  Gaussian approximation is required to understand if the peak is  a real physical effect. Such analysis is beyond the scope of the current paper.  

Our results are important for a number of reasons. 
For example, they may play a role in obtaining a more accurate and efficient numerical solution of  Eliashberg's equations  in the presence of strong Coulomb repulsion. Thus, it will be interesting to explore whether it can bring any advantage to {\it ab initio} techniques applied to Eliashberg theory~\cite{margine2013anisotropic,Aperis2015,Bekaert2016}. 

Regarding the physical implications of our theory, we have made concrete experimental predictions for the dependence of $H_{c2}$ on the repulsion strength. In particular, we predicted that when the Coulomb repulsion strength is tuned, an exceptional point in the fluctuation matrix can be manipulated to cause the slope of  $H_{c2}$ near $T_c$ to strongly peak. Such a prediction can be tested in experiments by looking at the thickness dependence of the upper critical field in thin films~\cite{zaytseva2020upper} or by directly controlling the screening of Coulomb repulsion in two-dimensional superconductors using  screening gates~\cite{liu2021tuning}. 

{
Our theory also applies to pairing interactions which compose of both attractive and repulsive  channels in momentum space. Examples include  the  Kohn-Luttinger mechanism~\cite{KohLuttinger} or systems with momentum-dependent orbital hybridization~\cite{kozii2015odd,kozii2019superconductivity}. 
In particular,  when space group symmetries are broken, different channels mix, thus coupling repulsive and attractive channels. 
Such symmetry breaking can come from the underlying lattice or from the expansion in momentum when considering collective modes.  Moreover, the non-linear form of the saddle-point equations implies that  the effect of repulsive channels cannot be neglected even when symmetries are conserved. Namely, the repulsive channels feed into the attractive ones, and vice versa, at non-linear order. We thus conclude that the existence of repulsive channels, which divert the saddle point into the complex plane,  is  a generic feature for both temporal and spatial decomposition of a realistic pairing interaction.

The eigenchannel decomposition picture also raises questions regarding the instability of a Fermi surfaces at zero temperature. According to Kohn-Luttinger theory} all Fermi surfaces are unstable to superconductivity at a sufficiently low temperature when time-reversal symmetry is present. However, in the case of the Morel-Anderson interaction, Eq.~\eqref{Eq:V_AM}, we find that repulsion can prevent an $s$-wave superconducting instability at zero temperature if the repulsion is strong enough. The reason for the absence of an instability is that the repulsive and attractive (frequency) channels are coupled [as shown by Eq.~\eqref{eq:gapFFbar}]. This raises the question whether the Kohn-Luttinger effect can be prevented, even at zero temperature, if all spatial
symmetries except for translations are broken such that repulsive and attractive (momentum) channels are mixed.

Finally, we conclude with a note. In a recent study the authors of Ref.~\cite{cohen2023complex} have shown that by extending the path integral of a frustrated spin ladder into a generalized complex plane they can significantly improve the  convergence of  determinant quantum Monte Carlo (DQMC) simulations with a sign problem. We find an interesting connection between this approach and ours, which may open a path to exact numerical simulation of superconductors with repulsion.

\section{Acknowledgements} We are grateful to Avraham Klein, Rafael Fernandes, Efrat Shimshoni, Udit Khanna, Andrey Chubukov, Amit Keren, Dimitri Pimenov, Herb Fertig, Ganapathy Murthy, and Mason Protter for helpful discussions. We are espcially grateful to Matan Ben-Dov for helping with the  gradient descent  method and to Grigory Tarnopolsky for insightful discussions about the saddle-point solution. J.R. also thanks Amit Keren for his invitation to give a lecture series in the Technion during which some of these ideas came to life. J.R.  acknowledges the
support of the Israeli Science Foundation under Grant No. 967/19.

\bibliography{ref}

\appendix

\section{Details of the numerical calculations}\label{app:numerics}

In this section we elaborate on the numerical calculations performed in this manuscript. The parameters used in the numerical calculations are summarized in Table~\ref{tab:Parameters}. 


\subsection{Saddle-point solution using the gradient descent method}
To obtain the saddle point within the ``gradient descent'' method we employ Eqs.~\eqref{eq:gradient passing eqs}, where we descend the action with respect to $f_{\eta}$.
The solution is obtained by guessing an initial ansatz, and then consequently updating it at each step according to Eqs.~\eqref{eq:gradient passing eqs}. We choose a uniform ansatz  $f_{\eta}=\bar{f}_\eta=0.2\omega_D$.
The procedure depends on the step size $e_\eta$. The most straightforward method would be to choose a fixed step size, e.g., $e_\eta=0.1$. However, a more efficient method is to choose the step size dynamically based on the second derivative using the Barzilai-Borwein  method  \cite{Barzilai1988Gradient}. 

The algorithm's stopping criterion was defined according to the mean difference between successive steps. Namely, we defined convergence to be the point where  the sum
\begin{align}
\label{app:StoppingCriteria}
    \sum_\w\left|\frac{\D^{i+1}(\w)-\D^{i}(\w)}{\D^{i}(\w)}\right|<0.001 \,,
\end{align}
i.e., dropped below  0.1\%.

Additionally, we truncate the number of eigenchannels used in the minimization procedure, $\eta < \eta_c$. As seen in Fig.~\ref{fig:EigLorentz} the weight of the eigenvlaues for the interaction in Eq.~\eqref{Eq:V_AM} decays exponentially with $\eta$, allowing us to obtain accurate results with only $\eta_c = 30$  channels.

\subsection{Solution of Eliashberg's gap equation using the iterative method}
The most straightforward way to solve the non-linear Eliashberg equation is by the method of iteration~\cite{margine2013anisotropic}. Using Eqs.~\eqref{eq:D1 and D2}, we can write a self-consistency equation of the form $\Delta(\omega)=F[\Delta(\omega')]$, where F is the non-linear Eliashberg operator. 
The iteration method is used by employing the relation 
\begin{align}
    \Delta^{i+1}(\omega)= F[\Delta^i(\omega)]
\end{align}
iteratively starting from a uniform ansatz $\D(\w)=0.2\omega_D$. By iterating the equation we typically obtain convergence quickly for small $\mu$, where convergence is determined by Eq.~\eqref{app:StoppingCriteria}.



\begin{table}
\caption{\label{tab:Parameters}
List of parameters used in different calculations}
\begin{ruledtabular}
\begin{tabular}{cccc}
 & Saddle-point solution (GL)  & Meaning \\
\hline
$\omega_D$ & $1(1)$ & Debye frequency \\
$\omega_c$ & $10\omega_D (30\omega_D)$ & Cutoff frequency \\
$\eta_c$ & 30(30) & Cutoff $\eta$ \\
$e_{\eta}$ & 0.1 & Step size for gradient descent
\end{tabular}
\end{ruledtabular}
\end{table}
\subsection{The frequency cutoff}
\label{app:cutoff}
The Matsubara sums performed in this paper were implemented in one of two manners. The first one is a ``hard'' high-frequency cutoff, i.e., $N=\lfloor{\omega_C}/{2\pi T}-1/2\rfloor$. This cutoff introduces a numerical complication that happens as we tune the temperature. Namely, as temperature is such that ${\omega_c}/{2\pi T}-1/2$ is an integer, the number of frequencies in the sum changes abruptly. This is the origin of the ``staircase'' artifact seen, for example, in Figs.~\ref{fig:eigs_fluctuation_matrix}(a) and (b). In order to mitigate this effect in the calculations of Figs.~\ref{fig:Hc2} and \ref{fig:Hc2HeatMap}, we used a ``soft'' cutoff, which smoothly reduces the weight of frequency tail. The weight of each frequency is then determined by 
\begin{align}
    n(\omega
    )=n_{FD}\left({\omega-\omega_c}\right),
\end{align}
where $n_{FD}(x)=(\exp \left[x/T_{eff}\right] +1)^{-1}$ is the Fermi-Dirac distribution, and $T_{eff}$ is the effective width taken to be $T_{eff}=10\pi T_c$.


\section{On the accuracy of gap calculations with a truncated interaction}
\label{sec:BCS}
The eigenvalue decomposition of the interaction introduces a natural scheme for a controlled approximate method that  speeds up numerical calculations. Namely, by truncating the number of eigen channels we keep in the calculation we can control the size of the matrices and the number of fields that need to be determined:
\[\left\{ f_\eta \right\}_{\eta = 1}^{N_\eta} \to \left\{ f_\eta \right\}_{\eta = 1}^{\eta_c}\,,\]
where $N_\eta$ denotes the full number of eigenvalues before truncation.
In this section we will explore the influence of such a truncation. 
To this end, we consider the interaction in Eq.~\eqref{Eq:V_AM}  in the limit where there is no repulsion ($\mu=0$):
\begin{align}
    \hat V(\omega-\omega')=-{\lambda\over N_F}\frac{\omega^2_D}{(\omega-\omega')^2+\omega^2_D}\,.
\end{align}

Because there are no repulsive eigenchannels in the interaction, one may write $\bar f_\eta = f_\eta^*$, and it is sufficient to solve the self-consistent equation for just one of the two fields:
\begin{equation}\label{eq:app:f}
{f_{\eta}\over \sqrt{T}}=
-\pi v_{\eta}\sum_{\omega}
\frac{\Delta_{\omega}U_{\omega,\eta}}
{\sqrt{\omega^2+|\Delta_{\omega}|^2}},
\end{equation}
where 
\[\Delta_{\w} \equiv \sqrt{T}\sum_{\eta=1}^{\eta_c} f_{\eta} U^*_{\eta,\w} \,,\]
and we have limited the number of fields to $\eta_c$.

\begin{figure*}
     \centering
\includegraphics[width=1\textwidth]{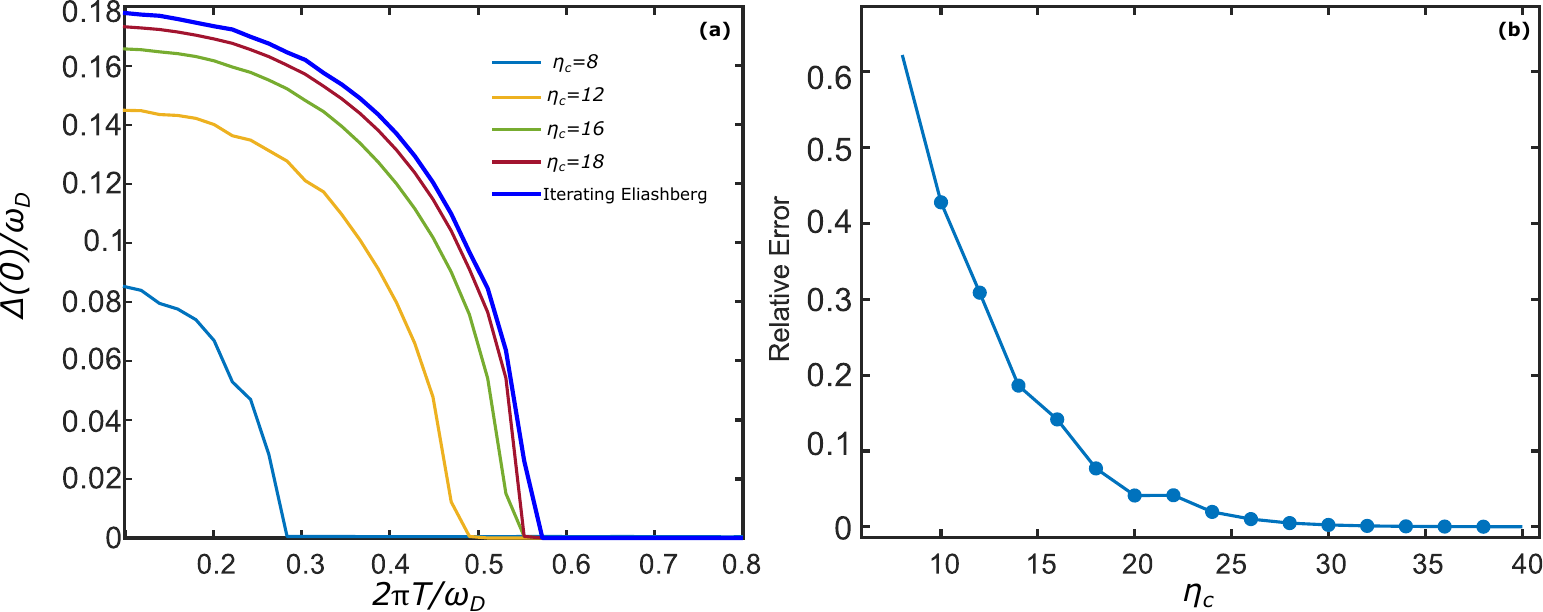}
         \caption{Eigenfunction method for a purely attractive BCS interaction $V(\omega-\omega')=-V_{0}{\omega^2_D}/{\left[(\omega-\omega')^2+\omega^2_D\right]}$. (a) The saddle-point solution, $\Delta(0)$, as a function of temperature, for different values of the eigenvalues cutoff $\eta_c$. Comparison between eigenvalue method and iterating Eliashberg solution.  (b) The relative error as a function of $\eta_c$.}
         \label{fig:Comparison}
\end{figure*}

The solution at $\w=0$ as a function of temperature for different cutoff values $\eta_c$ with $\lambda=0.67$ is shown in Fig.~\ref{fig:Comparison}~(a). These plots are compared with a  straightforward solution of the Eliashberg equation. 
We find that the solution of Eq.~\eqref{eq:app:f}  continuously converges to the Eliashberg solution as $\eta_c$ is increased. 

In Fig.~\ref{fig:Comparison}~(b) we plot the relative error between these two solutions as a function of $\eta_c$. The error decays exponentially and falls below $5\%$ at $\eta_c \approx 20$. We can understand this behavior by noting that 
the eigenvalues of the interaction $v_\eta$ decay exponentially at the same typical scale (see for example Fig.~\ref{fig:EigLorentz}). Thus, we conclude that when the eigenvalues decay exponentially with $\eta$ we can use the truncation method to accelerate calculations.  

\section{Computation of the fluctuation matrix}
\label{app:FulctMatrix}
Here we elaborate on the expansion of the action in Eq.~\eqref{eq:GL_action} in fluctuations about the saddle point (Section~\ref{sec:GL theory}):
\[f_\eta(Q) = f_\eta ^{(0)}+a_\eta(Q) + i b_\eta(Q)\,,\]
where $a_\eta$ and $b_\eta$  parametrize the fluctuations and, in general, can be complex themselves. 
In this paper we consider the expansion above $T_c$, so we set $f_\eta^{(0)}=0$. The quadratic term of the action \eqref{eq:GL_action} gives 
\begin{align*}
\frac{\bar{f}_{\eta}(Q) f_{\eta}(Q)}{|v_{\eta}|}=&\frac{a_{\eta}^2(Q)+b_{\eta}^2(Q)}{|v_{\eta}|}\,,
\end{align*}
while the $\mrm{tr}\ln$ part is expanded as follows 
{
\begin{widetext}
\begin{align}\label{eq:trlog expansion}
\tr{\ln{\mc G_k^{-1}(Q)}} \approx \tr{\ln{\hat{\mathds G}_k^{-1}}}-\frac{1}{2}\tr{\hat{\mathds G}_k\hat{\Delta}_{k}(Q)\hat{\mathds G}_{k+Q}\hat{\Delta}_{k+Q}}(-Q) =
    S_0 + G_0(-k-Q)G_0(k)\Delta_{1,k}(Q)\bar\Delta_{2,k}(Q)\,,
\end{align}
\end{widetext}
where we defined 
\begin{align}
        \hat{\mathds G}_k=\begin{pmatrix}
        G_0(k) & 0 \\
        0 & -G_0(-k)
    \end{pmatrix}
\end{align}
and 
\begin{align}\hat{\Delta}_k(Q)=\begin{pmatrix}
        0 & \Delta_{1,k}(Q) \\
        \bar\Delta_{2,k+Q}(-Q) & 0
    \end{pmatrix},
\end{align}
{such that $\mc G_k^{-1}(Q) = \hat{\mathds G}_k^{-1} \delta_{Q,0} - \hat{\Delta}_k(Q).$} The relation between $\Delta_{i,k}(Q)$ and fields $a_\eta(Q)$, $b_\eta(Q)$ is given by Eq.~\eqref{Eq:deltaab} with $\Delta_{i,k}^{(0)}=0$, and summation over frequencies/momenta $k$ and $Q$ in Eq.~\eqref{eq:trlog expansion} is implied. Note that normal-state self-energy corrections discussed in Sec.~\ref{sec:SE} can be straightforwardly incorporated by adding them to the bare Green's functions $G_0$.

}

The resulting action to second order in the fluctuations then assumes the form
\begin{equation}   \label{AppEq:S_2_naive}
    \mc S_{GL}^{(2)}= \sum_{Q}\vec{\alpha}^T(Q)
   \hat M(Q)
    \vec{\alpha}(Q)\,,
\end{equation}
where 
\begin{align}
\hat M(Q) = \hat{V}^{-1}-\hat{S}(Q) 
\end{align}
is the fluctuation matrix and $\vec{\alpha}^T(Q) = [a_\eta(Q),b_\eta(Q)]$ is a vector with dimension $2N_{\eta}$, where $N_\eta$ is the number of eigenchannels in Eq.~\eqref{Eq:S_I eigenbasis form}.
The matrices \[[\hat{V}^{-1}]_{\eta,\eta'} = |v_\eta|^{-1} \delta_{\eta, \eta'} \begin{pmatrix}
       1& 0 \\
       0  & 1 
         \end{pmatrix}\] 
and
\begin{align}\label{eq:S_matrix_app}
  \hat{S}_{\eta \eta'}(Q)=B_{\eta,\eta'}(Q)
    \begin{pmatrix}
       1& -i \\
       i  & 1 
         \end{pmatrix}
\end{align}
are $2N_\eta \times 2N_\eta$ matrices, where
\begin{align}\label{SMEq:Betaeta'}
B_{\eta,\eta'}(Q)=\zeta_{\eta}\zeta_{\eta'}\frac{T}{L^3}\sum_{k}U^*_{\eta,k}G_0(k)G_0(-k-Q)U_{\eta',k}.
\end{align}

Finally, we note that the quadratic form in Eq.~\eqref{AppEq:S_2_naive} is symmetric, consequently, only the symmetric part of $\hat M(Q)$ contributes:
\[\bs \a^T(Q)\hat M(Q)\bs \a(Q) =\bs \a^T(Q)\hat M_s(Q)\bs \a(Q)\,, \]
where (assuming that $U^*_{\eta,k} = U_{\eta,k}$)
\begin{align} 
\hat M_s(Q) &= \frac{\hat M(Q)+\hat M^T(Q)}2  \\ &=\left( \frac{\delta_{\eta,\eta'}}{|v_\eta|} - B_{\eta,\eta'}(Q)\right)  \begin{pmatrix} 1& 0 \\
       0  & 1 
         \end{pmatrix}. \nonumber
\end{align} 
\\

\subsection{Analytic calculation of pairing susceptibility}
\label{app:PairingSucept}
Here we compute the matrix elements $B_{\eta,\eta'}(|\bs Q|)$ defined in Eq.~\eqref{SMEq:Betaeta'}. We only consider the dependence of the matrix elements on momentum $|\bs Q|$ because we focus on the time-independent (i.e., zero-frequency) Ginzburg-Landau theory in  Eq.~\eqref{eq:free energy}.
In this case the superconducting susceptibility is given by  
\begin{widetext}
{
\begin{align}
    \chi_\w(|\bs Q|)=\frac{1}{L^3}\sum_{\bk}G_0(k)G_0(-k- Q)=\int \frac{d^3k}{(2\pi)^3} \frac{1}{-i\omega+\xi_\bk}\frac{1}{i\omega+\xi_{-\bk-\bs Q}},
\end{align}
where we used that $Q = \{0, \bs Q\}$. Assuming that $\xi_{-\bk} = \xi_{\bk}$ and $|\bs Q| \ll |\bk|\approx k_F$, where $k_F$ is the Fermi momentum, we can expand $\xi_{\bk + \bs Q} \approx \xi_\bk + v_F |\bs Q| \cos \theta,$ where $v_F$ is the Fermi velocity and $\theta$ is the angle between vectors $\bk$ and $\bs Q$. Then, changing integration variable from $\bk$ to $\xi_\bk$ and $\theta$, we obtain

\begin{align}
\chi_\w(|\bs Q|) &\approx \frac{N_F}2  \int_{-1}^1 dx  \int_{-\infty}^{\infty} d\xi \frac1{-i\omega + \xi} \cdot \frac1{i\omega + \xi + v_F |\bs Q| x} = \frac{i N_F \pi \,  \text{sign}(\omega)}{v_F |\bs Q|} \int_{-1}^{1} \frac{dx}{x+\frac{2i\omega}{v_F |\bs Q|}} \nonumber \\ &= \frac{i N_F \pi\,  \text{sign}(\omega)}{v_F |\bs Q|} \ln \frac{2 i \omega + v_F |\bs Q|}{2 i \omega - v_F |\bs Q|} = \frac{2 \pi N_F}{v_F |\bs Q|} \arctan \left( \frac{v_F |\bs Q|}{2|\omega|} \right),
\end{align}
where $N_F$ is the density of states per spin at the Fermi level, and we introduced a new integration variable $x = \cos \theta$. Since our interaction depends on frequency only and we are interested in the static limit, this expression can further be used in Eq.~\eqref{SMEq:Betaeta'}:
\begin{equation}
    B_{\eta,\eta'}(|\bs Q|)=\zeta_{\eta}\zeta_{\eta'}{T}\sum_{\w}U^*_{\eta,\w}U_{\eta',\w}\chi_\w(|\bs Q|).
\end{equation}
}\end{widetext}

\subsection{Toy model}
\label{app:ToyModel}
In the main text we made use of a simplified model with the interaction
\begin{align}   \hat V_{\omega,\omega'}=\frac{\lambda}{N_F}\left[\mu-\frac{1}{1+(\omega/\omega_D)^2}\frac{1}{1+(\omega'/\omega_D)^2}\right].
\end{align}
The advantage of this interaction is that it has only two  non-zero eigenvalues, one of which is positive and one is negative. Moreover, the interaction can be diagonalized analytically for any value of the cutoff.
To demonstrate how the decomposition works, we can write the interaction in the following basis
\begin{align}
    \hat{V}=\frac{\lambda}{N_F}\left(\mu |\mathbb{1} \rangle \langle \mathbb{1}| -|\ell\rangle\langle \ell| \right),
\end{align}
where $|\mathbb{1} \rangle$ is a vector with all its components equal to 1 and $|\ell \rangle$ is a Lorentzian vector in $\omega$ space, i.e., $\langle \omega |\ell \rangle ={1}/{(1+\omega^2/\omega_D^2)}$.\footnote{Here $\{|\w\rangle\}_{\w = -{\w_c}}^{\w_c}$ denotes an orthonormal set of states which are local in Matsubara space, such that $\langle \w' |\w\rangle = \d_{\w,\w'}$.} To construct an orthonormal basis we employ the Gram-Schmidt process:
\begin{align}
\begin{split}  
    &|\psi_1\rangle=\frac{1}{\sqrt{N_\omega}}|\mathbb{1}\rangle,\\
    &|\psi_2\rangle=  {1\over \sqrt A}\left(|\ell\rangle - |\psi_1 \rangle\langle \psi_1 |\ell\rangle \right),
\end{split}
\end{align}
where $N_\omega={\omega_c}/{\pi T}$  is the number of Matsubara frequencies bellow the cutoff, 
and $A =\langle \ell| \ell\rangle-|\langle \ell| \psi_1 \rangle|^2$ is a factor that ensures proper normalization of $|\psi_2\rangle$, $\langle \psi_2 | \psi_2\rangle = 1$.
Additionally, we define the projection
\begin{equation}
\langle\psi_1|\ell\rangle=\frac{1}{\sqrt{N_\omega}}\sum_{\omega = -\omega_c}^{\omega_c}\frac{1}{1+(\omega/\omega_D)^2}\equiv \sqrt{N_\ell}.
\end{equation}

In the limit $T \ll \omega_D \ll \omega_c$ that we consider in this paper, the sum can be replaced by an integral, and we find
\begin{align}
&\langle\psi_1|\ell\rangle \approx {1\over 2\pi T \sqrt{N_\omega}}\int_{-\infty} ^{\infty} {d\w \over 1+ \w^2 /\w_D^2}= {\w_D \over 2T \sqrt{N_\w}},\nonumber  \\
&\langle\ell|\ell\rangle \approx {1\over 2\pi T} \int_{-\infty}^\infty{d\w \over (1+\w^2/\w_D^2)^2} = {\w_D \over 4T},
\end{align}
such that $A \approx(\w_D/4T)(1-2/\sqrt{N_\w}) \approx (\w_D/4T)$.

In the general case, we can write the interaction in the basis of $|\psi_1\rangle$  and $|\psi_2\rangle$: 
\begin{align}\label{eq:app:int}
    \hat{V}=\frac{\lambda}{N_F}\begin{pmatrix}
        N_\omega \mu -N_\ell &- \sqrt{AN_\ell }\\
        -\sqrt{AN_\ell} & -A
        \end{pmatrix}.
\end{align}
\begin{widetext}

Diagonalizing the matrix in Eq.~\eqref{eq:app:int} we obtain the eigenvalues 
\begin{align}
    v_{\pm}=\frac{\lambda}{N_F} \left[\frac{N_\omega\mu-N_\ell -A}{2}\pm \sqrt{\frac{(N_\omega\mu-N_\ell+ A)^2}{4}+N_\ell A} \right]
\end{align}
and eigenvectors
\begin{align}
\label{app:ToyU1}
    \begin{matrix}
    U_+=\begin{pmatrix} 
        \cos {\theta}/{2}\\
        \sin {\theta}/{2}
    \end{pmatrix},    & & U_-=\begin{pmatrix} 
        -\sin {\theta}/{2}\\
        \cos {\theta}/{2}
    \end{pmatrix},    
    \end{matrix}
\end{align}
where we define 
\begin{align}
\label{app:ToyU2}
    \begin{split}
     & \cos{\theta}=\frac{N_\omega\mu-N_\ell+A}{\sqrt{(N_\omega\mu-N_\ell+ A)^2+4N_\ell A}}, \\   
     & \sin{\theta}=-\frac{2\sqrt{N_\ell A}}{\sqrt{(N_\omega\mu-N_\ell+A)^2+4N_\ell A}}.   
    \end{split}
\end{align}
The eigenvectors $U_{\pm}$ are plotted in Fig.~\ref{fig:ToyU}.

\begin{figure}[h]
     \centering
\includegraphics[width=0.5\textwidth]{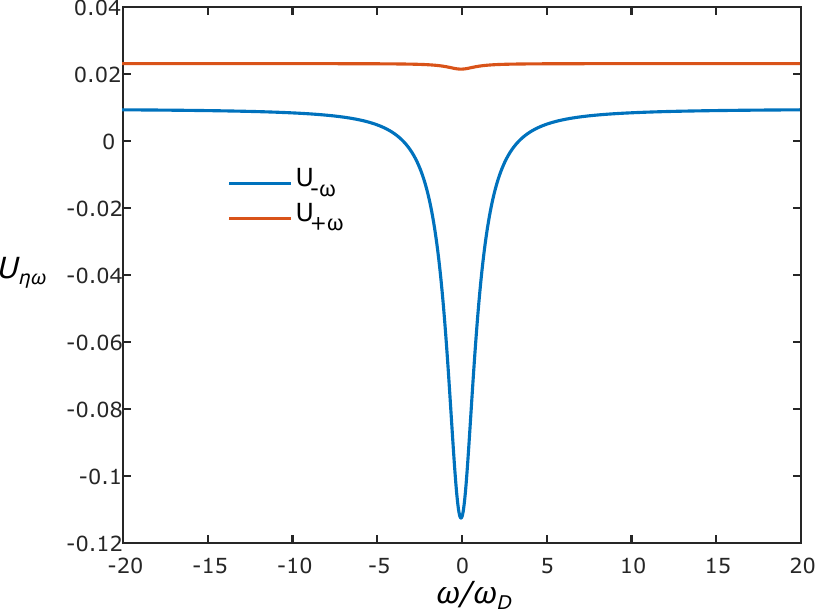}
         \caption{ The toy model interaction's eigenvectors, Eqs.~\eqref{app:ToyU1} and~\eqref{app:ToyU2},  for $\lambda=1$, $\mu=1$, and $\omega_c=30\omega_D$.}
         \label{fig:ToyU}
\end{figure}

Now we turn to derive the GL theory for this model (as described in Section~\ref{sec:GL theory}). The symmetrized fluctuation matrix assumes the form
\begin{align}
    \hat{M}_s=\begin{pmatrix}
        m_{-} & 0 & im_{\pm} & 0\\
        0 & m_{-} & 0 & im_{\pm}\\
        im_{\pm} & 0 & m_{+} & 0\\
        0 & im_{\pm} & 0 & m_{+}
    \end{pmatrix},
\end{align}
where the diagonal terms are 
\[m_-(|\bs Q|)=|v_{-}|^{-1}-{T}\sum_{\w}U^*_{-,\w}U_{-,\w}\chi_\w(|\bs Q|),\]
\[m_+(|\bs Q|)=|v_{+}|^{-1}+{T}\sum_{\w}U^*_{+,\w}U_{+,\w} \chi_\w(|\bs Q|),\]
and the off-diagonal elements are\[m_{\pm}(|\bs Q|)=-{T}\sum_{\w}U^*_{-,\omega}U_{+,\omega}\chi_\w(|\bs Q|).\]
Diagonalizing this matrix analytically results in two distinct doubly degenerate eigenvalues:
\begin{align}
   \hat \varepsilon_{1,2}=\frac{m_-+m_+}{2}\pm\frac{\sqrt{(m_- - m_+)^2-4m_\pm^2}}{2}.
\end{align}
The exceptional points are given by the condition $(m_--m_+)^2 = 4m_{\pm}^2$, where the two eigenvalues coalesce and the square root changes its value from purely real to purely imaginary. As a result, the matrix becomes non-diagonalizable at these points. In the region $(m_--m_+)^2 < 4m_{\pm}^2$, the two eigenvalues are complex conjugates of each other, $\hat \varepsilon_1 = \hat \varepsilon_2^*$, so their absolute values are equal, $\varepsilon_1 = |\hat \varepsilon_1| = |\hat \varepsilon_2| = \varepsilon_2$, as is shown in Fig.~\ref{fig:eigs_fluctuation_matrix}.

\end{widetext}

\end{document}